\documentclass[authoryear,3p,twocolumn,nofootinbib]{emulateapj-rtx4}
\usepackage[english]{babel}
\usepackage{graphicx}
\usepackage{dcolumn}
\usepackage{bm}
\usepackage{longtable}
\usepackage{amsmath}
\usepackage{amssymb}
\usepackage{latexsym}
\usepackage{mathrsfs}
\usepackage{amsfonts}
\usepackage[usenames]{color}
\usepackage{syntonly}
\usepackage{natbib}
\usepackage{epstopdf}
\usepackage{rotating}
\usepackage{latexsym}
\usepackage{verbatim}
\usepackage{hyperref}
\usepackage{fancyheadings}
\usepackage{fancyhdr}

\newcommand{\be} {\begin{equation}}
\newcommand{\ee} {\end{equation}}
\newcommand{\de} {\,\mathrm{d}}
\newcommand{\bea} {\begin{eqnarray}}
\newcommand{\eea} {\end{eqnarray}}
\newcommand{\bdm} {\begin{displaymath}}
\newcommand{\edm} {\end{displaymath}}
\newcommand{\ba} {\begin{array}}
\newcommand{\ea} {\end{array}}

\newcommand{\bfg}  {\begin{figure}}
\newcommand{\efg}  {\end{figure}}
\newcommand{\bfgd}  {\begin{figure*}}
\newcommand{\efgd}  {\end{figure*}}
\newcommand{\incgr} {\includegraphics}

\newcommand{\btb} {\begin{table}}
\newcommand{\etb} {\end{table}}
\newcommand{\ben} {\begin{enumerate}}
\newcommand{\een} {\end{enumerate}}

\newcommand{\bit} {\begin{itemize}}
\newcommand{\eit} {\end{itemize}}

\newcommand{\wmap} {{\slshape WMAP~}}

\newcommand{\healpix} {{\slshape HEALPix~}}

\newcommand{\fnl}{f_{\rm NL}}

\def\aap  {A\&A}

\def\apjl {ApJ}
\def\apjs {ApJ}

\begin{document}

\newcommand{\hg}{{\hat\gamma}}

\bibliographystyle{apsrev}

\title{NeedATool: a Needlet Analysis Tool for cosmological data processing}

\author{Davide Pietrobon}
\email{davide.pietrobon@jpl.nasa.gov}
\affiliation{Jet Propulsion Laboratory, California Institute of Technology\\
4800 Oak Grove Dr.~ 91109 Pasadena CA}

\author{Amedeo Balbi}
\email{amedeo.balbi@roma2.infn.it}
\affiliation{Dipartimento di Fisica, Universit\`a di Roma ``Tor Vergata'',
via della Ricerca Scientifica 1, 00133 Roma, Italy\\INFN Sezione di
Roma ``Tor Vergata'', via della Ricerca Scientifica 1, 00133 Roma, Italy}

\author{Paolo Cabella}
\email{paolo.cabella@roma2.infn.it}
\affiliation{Dipartimento di Fisica, Universit\`a di Roma ``Tor Vergata'',
via della Ricerca Scientifica 1, 00133 Roma, Italy}

\author{Krzysztof M.~  Gorski}
\email{krzysztof.m.gorski@jpl.nasa.gov}
\affiliation{Jet Propulsion Laboratory, California Institute of Technology\\
4800 Oak Grove Dr.~ 91109 Pasadena CA\\
Warsaw University Observatory, Aleje Ujazdowskie 4, 00478 Warszawa, Poland}
%
\begin{abstract}
We introduce NeedATool (Needlet Analysis Tool), a software for data analysis based on needlets, a wavelet rendition which is powerful for the analysis of fields defined on a sphere. Needlets have been applied successfully to the treatment of astrophysical and cosmological observations, and in particular to the analysis of cosmic microwave background (CMB) data. 

Usually, such analyses are performed in real space as well as in its dual domain, the harmonic one. Both spaces have advantages and
disadvantages: for example, in pixel space it is easier to deal with partial sky coverage and experimental noise; in harmonic domain, beam treatment and comparison with theoretical predictions are more effective. During the last decade, however, wavelets have emerged as a useful tool for CMB data analysis, since they allow to combine most of the advantages of the two spaces, one of the main reasons being their sharp localisation. 

In this paper, we outline the analytical properties of needlets and discuss the main features of the numerical code, which should be a valuable addition to the CMB analyst's toolbox.
\end{abstract}



\maketitle

Over the last two decades, the detailed analysis
of cosmic microwave background radiation anisotropies has been
fundamental in determining the global properties of our Universe and its
evolutions. The cosmological concordance model encodes into few parameters the variety of processes we observe in the local universe as well as those occurring at very large scales. Such parameters
have been measured very precisely by several CMB experiments \citep{Mather:1992,Smoot:1992,deBernardis:2000gy,Komatsu:2010wmap7}. and such measurements will be further refined with the next generation of
cosmological experiments. CMB data analysis is very demanding -- both in terms of computational power required and sophistication of the necessary techniques -- due to the complexity of the datasets and to the high degree of accuracy one wants to achieve. An overview on the techniques recently applied to astrophysical data analysis can be found in \citet{Pesenson:2010}. This holds in particular for cutting-edge experiments such as the ongoing Planck satellite \footnote{http://www.rssd.esa.int/SA/PLANCK/docs/Bluebook-ESA-SCI(2005)1.pdf}.

Over the last decade, wavelets \citep{Freeden:1998,AntoineVandergheynst1999,McEwen2006a,McEwen2007,Sanz2006, Starck2005,Starck:2009a} have emerged as a very powerful tool for CMB data analysis; a very incomplete list of references should include testing for non-Gaussianity \citep{Vielva2004NG,Cabella2004}, foreground subtraction \citep{Hansen2006}, point source detection \citep{Sanz2006}, component separation \citep{Moudden:2004wi,Starck2005}, polarisation analysis \citep{CabellaNatoliSilk2007} and many others. The reason for such a
strong interest is easily understood. As it is well-known, CMB models are best analysed in the frequency domain, where the behaviour at different
multipoles can be investigated separately; on the other hand, partial sky
coverage and other missing observations make the evaluation of
exact spherical harmonic transforms troublesome. The combination of these two features makes the time-frequency localisation properties of wavelets most valuable. See \citet{Starck:2009b} for a recent review on this topic\footnote{A mutliresolution package for data analysis and compression is also available at this url: http://jstarck.free.fr/mresol.htm} and \citet{Wiaux:2008} and for a numerical implementation.


Recently, a novel approach to spherical wavelets has been introduced in the statistical literature by \citet{Baldi2006}, adapting tools proposed in the functional analysis literature by \citet{NarcowichPetrushevWard2006};
the first application to CMB data is due to \citet{Pietrobon2006ISW}, where needlets are used to estimate (cross-)angular power spectra in order to search for dark energy imprints on the correlation between large scale structures and the CMB \citep{SachsWolfe1967}. \citet{Guilloux:2007} investigate the effect of different window functions in needlets constructions; whereas \citet{Baldi2007} provide further mathematical results on their behaviour for partially observed sky-maps. Needlets have been applied to angular power spectrum estimation in the presence of noise \citep{Fay2008,Fay2008b}, as well as to the estimation of the bispectrum \citep{Lan2008NeedBis}; the latter tool has been applied to the WMAP 5-year data release by \citet{Pietrobon2008NG} and \citet{Rudjord2009needBis} to constrain the primordial non-Gaussianity parameter. The bispectrum formalism has been further exploited by \citet{Pietrobon:2009qg} and \citet{Rudjord:2009au}, who addressed the sky asymmetry issue analysing respectively the properties of 3-point correlation function, and the primordial non-Gaussianity parameter. \citet{Cabella:2009} developed the bispectrum estimator including the marginalisation over the possible foreground residuals in the CMB maps, while \citet{Delabrouille2008maps,Ghosh:2010} produced a foreground component separation algorithm. The analysis of directional data are described in
\citet{Baldi2008Adaptivedensity}. Finally, the needlet formalism has been extended to the the polarisation field, as discussed by \citet{Geller2008Mat,Geller:2008a,Geller:2009a,Geller2008SpinNee,Geller:2009Spin,Geller:2010b}.

The aim of this paper is to describe a numerical code, called NeedATool (Needlet Analysis Tool), which has already been used in several of the above mentioned analyses. We first provide a discussion of the needlet formalism in Sec.~\ref{sec:implementation} and \ref{sec:formalism}. We then describe a viable implementation of the code in Sec.~\ref{sec:code} and we add our concluding remarks in Sec.~\ref{sec:conclusions}.

\section{Needlets Frame}
\label{sec:implementation}
Needlets enjoy several features which are not in general granted by other spherical wavelets construction.
Here, we recall some of these properties and refer to \citet{Marinucci2008,Lan2008NeedBis} for a comprehensive mathematical discussion. Complementary mathematical analyses can be found in \citet{Geller2007,Lan2008,Mayeli2008,Geller:2008a,Geller:2009a}. In particular, needlets
\bit
\item[a)] do not rely on any tangent plane approximation (compare \citealt{Sanz2006}), and take advantage of the manifold structure of the sphere;

\item[b)] being defined in harmonic space, they are computationally very convenient, and natively adapted to
standard packages such as \healpix\footnote{http://healpix.jpl.nasa.gov} \citep{Gorski2005Healpix};

\item[c)] they allow for a simple reconstruction formula (see Eq.~\ref{eq:recfor}),
where the same needlets functions appear both in the direct and the inverse
transform (see also \cite{Kerk2007});

\item[d)] they are quasi-exponentially (i.e. faster than any polynomial) concentrated in pixel space, see Eq.~\ref{eq:expine} below;

\item[e)] they are exactly localised on a finite number of multipoles; the
width of this support is explicitly known, controlling the power
encoded in each multipole range (see Eq.~\ref{eq:needlets_expansion});

\item[f)] needlets coefficients can be shown to be asymptotically uncorrelated (and hence, in the Gaussian case, independent) at any fixed angular distance, when the frequency increases.
\eit

\begin{figure}[h]
\begin{center}
\includegraphics[width=.8\columnwidth]{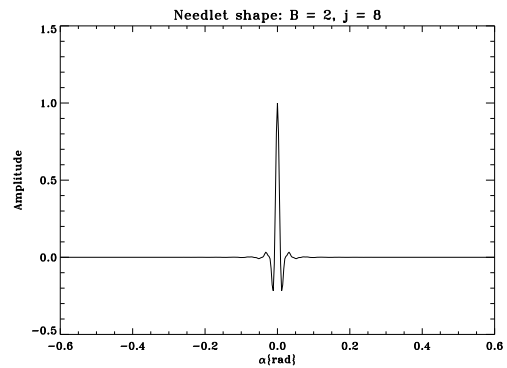}
\caption[Needlets in pixel space]{\small{Needlets in pixel
    space. $B=2$, $j=8$ as a function of the angle between
    $\hat\gamma$ and $\xi_{\rm jk}$}}
\label{fig:needdir}
\end{center}
\end{figure}

We first recall that the spherical needlet function is defined as 
\begin{equation}
\psi_{jk}(\hat\gamma) =\sqrt{\lambda_{jk}}\sum_{\ell}b(\frac{\ell}{B^{j}})\sum_{m=-\ell}^{\ell}\overline{Y}%
_{\ell m}(\hat\gamma)Y_{\ell m}(\xi _{jk});
\label{eq:needlets_expansion}
\end{equation}
where $\gamma$ and $\xi_{\rm jk}$ are directions on the sphere, $Y_{\ell m}$ is a spherical harmonic function, with $\overline{Y}_{\ell m}$ identifying its complex conjugate, and $b(x)$ is a filter function defined for $x\in[1/B,B]$, which the entire needlet construction relies on. Here, we use $\left\{ \xi _{jk}\right\} $ to denote a set of \emph{cubature
points} on the sphere, corresponding to frequency $j;$ and
$\lambda_{jk}$ denotes the cubature weights. In Fig.~\ref{fig:needdir} the needlet profile as function of the angle between $\hat\gamma$ and $\xi_{\rm jk}$ is shown for the choice B=2, j=8.

Intuitively, needlets should be viewed as a convolution of the projection
operator $\sum_{m=-\ell}^{\ell}\overline{Y}_{\ell m}(\hat\gamma)Y_{\ell m}(\xi _{jk})$  with a suitably chosen window function 
$b(\cdot)$. The needlet frame construction strongly relies on the spherical harmonic decomposition which represents the mathematical environment for the derivation of the fundamental properties of needlets \citep{Lan2008}: in particular, the existence of a reconstruction formula, as first pointed out by \citet{Freeden:1997,Freeden:1998}. More details can be found in \citet{Freeden:2002}. A similar approach has been followed by \citet{Starck2005,Starck:2009a}, who implemented wavelets, ridgelets and curvelets built directly on the sphere, both for scalar and spin-2 fields. All these sets of functions are examples of \emph{frames} on the sphere: they are over-complete and redundant but admit a well defined backward transformation. For this reason, frames are not only suitable for a multi-frequency analysis of a signal, as any other wavelet implementation, but also for data compression, denoising algorithm and component separation, as described by \citet{Moudden:2004wi,Starck:2009b}.

Besides spherical harmonic decomposition, needlet properties strongly depend on the the filter function $b(\cdot)$, which controls the angular scale span covered by each needlet and ensures that needlets enjoy quasi-exponential localisation properties in pixel space. Formally, we must ensure
that:

\begin{itemize}
\item[  i)] $b^{2}(\cdot)$ has support in $[\frac{1}{B},B],$ and hence $b(\frac{\ell}{%
B^{j}})$ has support in $\ell\in \lbrack B^{j-1},B^{j+1}]$

\item[ ii)] the function $b(\cdot)$ is infinitely differentiable in $(0,\infty ).$

\item[iii)] we have%
\begin{equation}
  \sum_{j=1}^{\infty }b^{2}(\frac{\ell}{B^{j}})\equiv 1\textrm{ for all }\ell>B.
  \label{eq:partun}
\end{equation}
\end{itemize}

It is immediate to see that property (i) ensures the needlets have bounded
support in the harmonic domain; property (ii) is the crucial element in the
derivation of the localisation properties \citep{NarcowichPetrushevWard2006}; finally, property (iii) is necessary to establish the
reconstruction formula (Eq.~\ref{eq:recfor}). Functions such as $b^{2}(\cdot)$ are called \emph{partitions of
unity}.

There are of course many possible constructions satisfying
(i-iii); indeed an interesting theme of research is the derivation
of optimal windows satisfying these three conditions (compare
\citealt{Guilloux:2007}), although the choice of $b(\cdot)$ is expected to
exert second-order effects on
the final estimates \citep{Lan2008}. An explicit recipe for the
construction of $b(\cdot)$ is given in Sec.~\ref{sec:code}. Very recently an extended study on how needlet properties depend on the filter functions has been conducted by \citet{Scodeller:2010}. Interestingly, the authors explore a peculiar construction able to mimic the Spherical Mexican Hat Wavelets based on mathematical study of \citet{Geller:2008a,Geller:2009a}.

Needlets coefficients are hence given by
\begin{eqnarray}
\beta _{jk} =\sqrt{\lambda_{\rm jk}}\sum_\ell b\left(\frac{\ell}{B^{j}}
\right)\sum_{m=-\ell}^{l}a_{\ell m}Y_{\ell m}(\xi _{jk})\text{.}
  \label{eq:needcoef}
\end{eqnarray}
In Fig.~\ref{fig:need_coef} we show the needlet
coefficients of WMAP 5-year temperature map for the specific choice
$B=2$ and $j=4$. A remarkable aspect of this construction is that the needlet coefficients can be represented easily as a Mollweide projection in the \healpix pixelization framework \citep{Gorski2005Healpix}, the most widely used tool for visualization and analysis of CMB maps. This makes dealing with needlets particularly handy, since it is easy to implement a needlet analysis code which exploits pre-existing \healpix routines.
\begin{figure}[h]
\center
\incgr[width=0.5\columnwidth, angle=90]{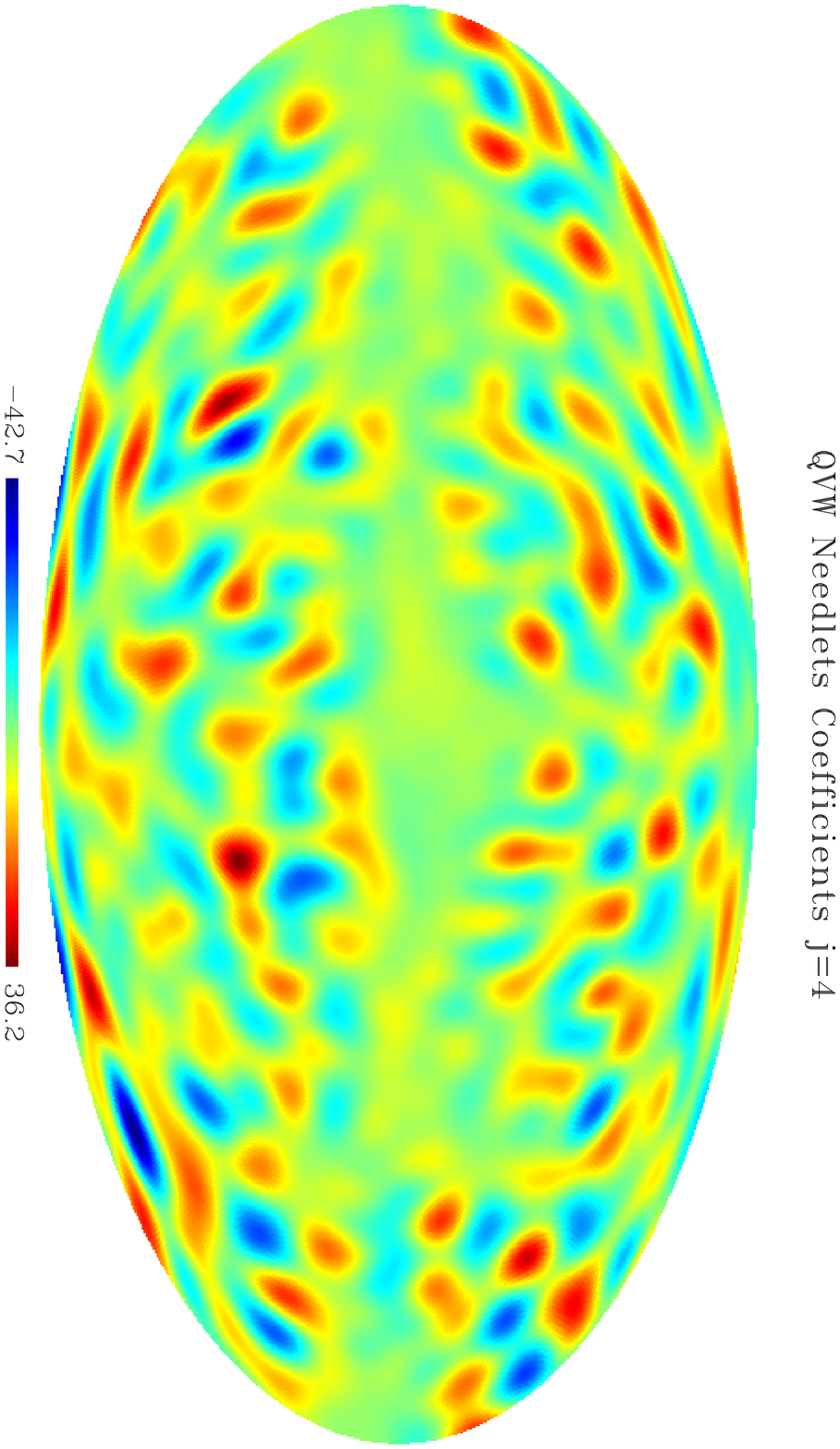}
\caption[Needlet coefficients]{\small{Needlet coefficients of the combined Q, V, W map at the
resolution $j=4$. The $B$ parameter is fixed to $2$. Notice as the anomalous bright spots found by \cite{Pietrobon2008AISO} are clearly visible.}}
\label{fig:need_coef}
\efg

It is very important to stress that, although the needlets do \emph{not} make up an orthonormal basis for square integrable functions on the sphere, they do represent a \emph{tight frame. }In general, a tight frame on the sphere is a countable set of functions $\left\{ e_{j}\right\} $ such that, for all square integrable functions on the sphere $f\in L^{2}(S^{2}),$ we have
\[
\sum_{j}\langle f,e_{j}\rangle^2\equiv \int_{S^{2}}f(\hat\gamma)^{2}d\Omega,
\]
so that the norm is preserved. Here $\langle f,g\rangle$ means the scalar product,
or more properly the projection between the functions $f$ and $g$. Of course, this norm-preserving property is
shared by all orthonormal systems; however, frames do not in general make up
a basis, as they admit redundant elements. They can be viewed as the closest
system to a basis, for a given redundancy, see
\citet{HernandezWeiss1996}, \citet{Baldi2006} and \citet{Baldi2007} for
further definitions and discussion. In our framework, the norm-preserving property translates into
\begin{equation}
\sum_{j,k}\beta _{jk}^{2}\equiv \sum_{\ell=1}^{\infty }\frac{2\ell+1}{4\pi }
\widehat{C}_{\ell}\textrm{ ,}
\label{eq:frame}
\end{equation}
where 
\be
\widehat{C}_\ell= \frac{4\pi}{2\ell+1}\sum_m |a_{\ell m}|^2 \nonumber
\ee
is the angular power spectrum of the map $T(\hat\gamma)$. Identity
\ref{eq:frame} has indeed been verified by means of numerical
simulations and implicitly provides the correct normalization for
needlets, $\lambda_{\rm jk}$. It is basically a consequence of the peculiar
partition-of-unity property of $b(\cdot)$ (Eq.~\ref{eq:partun}). Eq.~\ref{eq:frame} is related to a much more fundamental result, i.e.
the reconstruction formula
\begin{equation}
T(\hat\gamma)\equiv \sum_{j,k}\beta _{jk}\psi _{jk}(\hat\gamma)\text{,}
\label{eq:recfor}
\end{equation}
which in turn is a non-trivial consequence of the analythical
properties of the $b_{\ell,j}$ functions. 
 As mentioned before, the simple reconstruction formula of Eq.~\ref{eq:recfor} is typical of tight frames but does not hold in general for other wavelets systems.


The following quasi-exponential localisation property of needlets is due to
\citet{NarcowichPetrushevWard2006} and motivates their name. For any $M=1,2,...$ there exists a positive constant $c_{M}$ such that for
any point $x\in S^{2}$ we have
\begin{equation}
  |\psi _{jk}(\hat\gamma)|\leq \frac{c_{M}B^{j}}{(1+B^{j}\arccos (\langle x,\xi _{jk}\rangle)^{M}}%
\textrm{ .}  
  \label{eq:expine}
\end{equation}
We recall that $\arccos (\langle x,\xi _{jk}\rangle)$ is just the natural distance on the
unit sphere between the points $(x,\xi _{jk}).$ The meaning of Eq.~\ref{eq:expine} is then clear: for any fixed angular distance, the value of $\psi _{jk}(\hat\gamma)$
goes to zero quasi-exponentially in the parameter $B.$  This clearly
establish an excellent localization behaviour in pixel space. Note
that the constants $c_{M}$ do depend on the form of the weight
function $b(\cdot)$, and in particular on the value of the bandwidth
parameter $B$; typically a better localisation in multipole space
(i.e., a value of $B$ very close to unity) will entail a larger value
of $c_{M}$, that is, less concentration in pixel space for any fixed
$j$. The resulting trade-off in
the behaviour over the harmonic and real spaces is expected: smaller values
of $B$ correspond to a tighter localisation in harmonic space (less
multipoles entering into any needlet), whereas larger values ensure a faster
decay in real space \citep{Lan2008}.

In \citet{Baldi2006},  another relevant
property of needlets coefficients was discussed, namely their asymptotic
uncorrelation at any fixed angular distance, for growing frequencies $j$. More
explicitly, at high frequency needlets coefficients can be approximated as a
sample of identically distributed and independent (under Gaussianity)
coefficients. Also, in view of Eq.~\ref{eq:needcoef}, for full sky maps and in the absence of
any mask we should expect the theoretical correlation to be identically zero
whenever $\left| j_{1}-j_{2}\right| \geq 2$. This has been indeed numerically verified by \citet{Marinucci2008}.

The probabilistic properties of the needlet coefficients $\beta _{jk}$ have been established in \citet{Baldi2007}; in that paper, it is shown that for any two (sequence of) pixels $\xi _{jk},\xi _{jk^{\prime }}$ such that their angular distance is larger than a positive $\varepsilon ,$ for all $j$, we have
\begin{equation}
\frac{\langle\beta _{jk}\beta _{jk^{\prime }}\rangle}{\sqrt{\langle\beta
_{jk}^{2}\rangle\langle\beta _{jk^{\prime }}^{2}}\rangle}\leq \frac{c_{M}}{(s^{j}\varepsilon )^{M-1}}\text{ for all }M=1,2,3,\dots
\label{eq:baldi}
\end{equation}
thus proving wavelets coefficients are asymptotically uncorrelated as
$j\rightarrow \infty $ for any fixed angular distance. 
Eq.~\ref{eq:baldi} can then be seen as the statistical counterpart of
Eq.~\ref{eq:expine}. 

These properties are the basis for the large success of needlets as a
CMB toolbox, in particular when dealing with masked datasets.

\section[2- and 3- point correlation
functions]{Needlets Estimators: 2- and 3- point correlation
functions}
\label{sec:formalism}

Having introduced the spherical needlet
frame, and recalled the main properties which make needlets perform
extremely well in a wide number of applications to 2-dimension fields on the sphere, we now briefly describe some important statistical techniques largely used in CMB data
analysis. 

\subsection{(Cross-) Power Spectrum}

After computing the needlets coefficients $\beta_{jk}$ from a
2-dimension map (e.~g.~ the CMB or source count map), we can use Eq.~\ref{eq:frame} to build a (cross-)correlation estimator in wavelet space, $\beta_j$, as:
\begin{equation}
\beta_j^{\rm IJ}\equiv\sum_k {1\over N_{\rm pix}(j)}\beta_{jk}^{\rm I}\beta_{jk}^{\rm J}\text{,}
\label{eq:betaj}
\end{equation}
where $N_{\rm pix}(j)$ is the number of pixels (e.~g.~ in the \healpix scheme $N_{\rm pix}=12 N_{\rm side}^2$) with I and J
denoting the two different maps.
The theoretical prediction for $\beta_j$ can be computed from the
expected $C_l^{\rm IJ}$ as:
\begin{equation}
\label{eq:betaj_th}
\beta_j^{\rm IJ}=\sum_\ell {(2\ell+1)\over
4\pi}\,\left[b\left(\frac{\ell}{B^{j}}\right)\right]^2C_\ell^{\rm
IJ}\text{,}
\end{equation}
where we recall $C_\ell^{\rm IJ}\equiv\langle a_{\ell m}^{\rm I}a_{\ell
m}^{\rm J*}\rangle$ is the (cross-) angular power spectrum.

$\beta_j$ provides then an unbiased estimator for the (cross-) angular power
spectrum within the needlets framework. The analytic relation between
$\beta_j$ and $C_\ell$ underlines few more advantages in
using needlets. Indeed, it makes extremely easy and
straightforward dealing with beam profiles and experimental window
functions, which have to be taken into account when analysing real
data (see \citet{Pietrobon2006ISW}). The duality which needlets embed, namely the localisation
both in pixel and harmonic domain, allows also to characterise the noise properties (see
\citet{Delabrouille2008maps} and \citet{Pietrobon2008NG,Pietrobon:2009qg} for direct
applications to \wmap 5-year data.)

Computing the 4-point correlation function, it can be easily shown that the analytical expression for the
dispersion of the estimated cross-correlation power spectrum in
needlet space is:
\begin{equation*}
\label{eq:betaj_err}
\Delta \beta_j^{\rm IJ} =\sqrt{\sum_\ell \frac{(2\ell+1)}{16\pi^{2}}
\left[b\left(\frac{\ell}{B^{j}}\right)\right]^4 \left(\left(C^{\rm
IJ}_\ell\right)^{2} + C^{\rm I}_\ell C^{\rm J}_\ell\right)}\text{,}
\end{equation*}
which, of course, must be only taken as an approximation when dealing
with real data, where window functions, noise and partial sky
coverage have to be taken into proper account.

It is important to stress that Eq.~\ref{eq:betaj} generalises into
\be
\beta_{j_1j_2}^{\rm IJ} = \frac{1}{N_{\rm pix}}\sum_k
\beta_{j_1k}^{\rm I}\beta_{j_2k}^{\rm J}\text{,}
\ee
which describes the needlets coefficients covariance and it has been
used in \citet{Pietrobon2008AISO} to determine the degree of anomaly of
a few hot and cold spots found in the CMB temperature map.

We have shown that the needlets formalism may be suitable for the
problem of angular power spectrum estimation from a CMB map (and therefore, indirectly, to the estimation of cosmological parameters). In particular the application of needlets to the \wmap
3-year data led to interesting constraints on the dynamics of dark
energy \citep{Pietrobon2006ISW} and to the measure of the difference in power between the two estimates of the power spectrum computed on the north and south CMB skies \citep{Pietrobon2008AISO}. A detailed discussion on the application of needlets to power spectrum estimation can be found in \citet{Fay2008}.

\subsection{Needlets Bispectrum}
\label{subsec:needBin}

In the previous section we described how needlets can naturally be
applied to the estimation of the 2-point correlation function and how,
thanks to the reconstruction formula (Eqs.~\ref{eq:partun} and \ref{eq:frame}), it relates to the usual angular power spectrum. It is easy to extend the formalism to the higher order correlation functions.

Here, we focus on the 3-point correlation function, which plays a crucial role in CMB data analysis to detect any departure of primordial fluctuations from the Gaussian statistics, a smoking gun for non-standard inflationary models. We next briefly review the properties of the needlet bispectrum and how it relates to the spherical harmonic bispectrum. An extensive discussion is provided in \citet{Lan2008NeedBis,Pietrobon2008NG,Rudjord2009needBis,Pietrobon:2009qg,Rudjord:2009au}.

The needlet bispectrum is defined as follows:
\bea
S_{j_1j_2j_3} &\equiv&\frac{1}{N_{\rm pix}}
\label{eq:needBis}
\sum_k \beta_{j_1k}\beta_{j_2k}\beta_{j_3k} \\
&=&
\sum_{\ell_1\ell_2\ell_3}b_{\ell_1}^{(j_1)}b_{\ell_2}^{(j_2)}b_{\ell_3}^{(j_3)}\nonumber
\\
&\times&\sqrt{\frac{(2\ell_1+1)(2\ell_2+1)(2\ell_3+1)}{4\pi}} \nonumber \\
&\times&\left(
\begin{array}{ccc}
\ell_1 & \ell_2 & \ell_3 \\
0 & 0 & 0 \\
\end{array}
\right)\hat{\rm B}_{\ell_1\ell_2\ell_3} \nonumber \text{,}
\eea
where 
\be
\hat{\rm B}_{\ell_1\ell_2\ell_3}\equiv\langle
a_{\ell_1m_1}a_{\ell_2m_2}a_{\ell_3m_3}\rangle = \sum_m a_{\ell_1m_1}a_{\ell_2m_2}a_{\ell_3m_3}\nonumber
\ee
is the estimated bispectrum, averaged over $m_i$s. $S_{j_1j_2j_3}$ can
be seen as a \emph{binned bispectrum}, a smooth and combined component
of the angular bispectrum. The bispectrum is supposed to be vanishing for a Gaussian
distribution. Standard inflation mechanism \citep{Guth1981,Sato1981,Linde1982,AlbrechtSteinhardt1982} predicts a tiny non-Gaussianity in the
cosmological perturbations: this is why a great effort has been spent to measure a
bispectrum amplitude different from zero in the CMB data, which would
provide an extraordinary handle on the early universe physics \citep{Smith2009fnlFore,Curto2008Archeops,Komatsu2008wmap5,Natoli:2009,Smidt:2009ir,Curto:2010}. This
kind of study is usually performed in terms of the non-linear
parameter $\fnl$ (see for example
\citet{KomatsuSpergel2001,Bartolo2004NGreview,SmithZaldarriaga2006}). 

One of the key properties of needlets, is that the sum of the
squared filter functions in harmonic space, $b_{\ell j}$, is 1 (see
Eq.~\ref{eq:partun}). This means that even if we group multipoles and each needlet peaks at a certain multipole range the total power is conserved. 
 Therefore, the
needlets power spectrum analysis can be in principle done with any
choice of the parameter $B$, being the total power conserved and only
the correlation and localisation properties affected by different
width of $b_{\ell j}$. We incidentally notice that any wavelet function defined as the difference between the square roots of the 
scaling functions at two different resolutions satisfies Eq.~\ref{eq:partun}, but the uncorrelation properties are in general not granted, since they are determined by the shape of the filter function $b_\ell(\cdot)$. See for example \citet{Scodeller:2010} for a detailed discussion on needlet filters.

This does not hold any more when the the cubic power of the filter
functions contributes to the estimator used, which is indeed
the case of the skewness expression and more
generally of the bispectrum one. This fact is displayed in
Fig.~\ref{fig:b_sum} where we plot the sum of square and cube of the
filters functions. The not uniform sampling of the multipoles for a
n-power estimator suggests that the choice of the B parameter is
crucial for the analysis and must be driven by
the insight on the range of multipoles to be probed.

\bfg[h]
\center
  \incgr[width=.8\columnwidth]{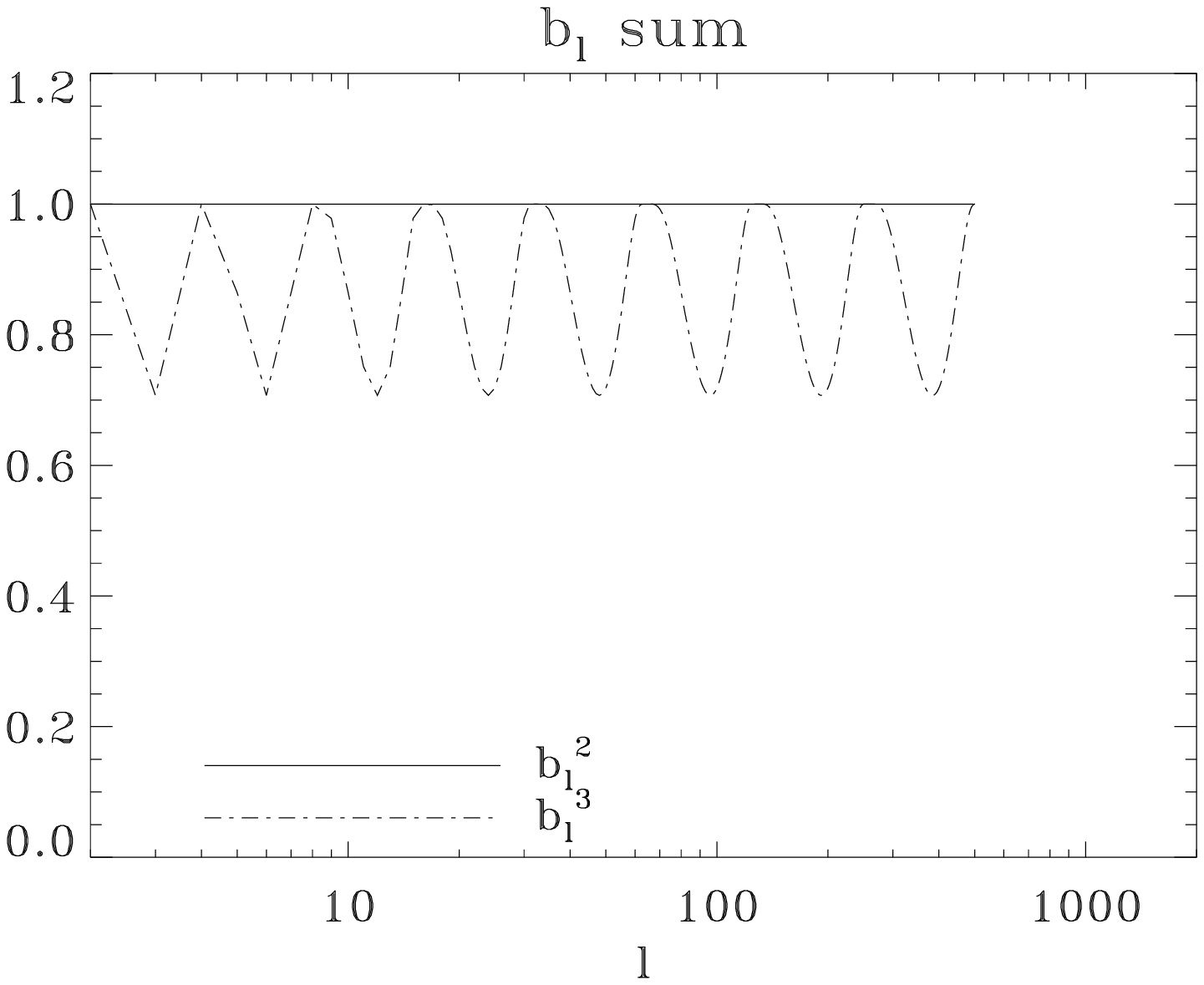}
  \caption[$b_\ell^2$ sum]{\small{Solid line sum of the $b_\ell^2$; dot-dashed line sum of the $b_\ell^3$. While the former is equal to 1 for the entire range of multipole, the latter is not.}}
  \label{fig:b_sum}
\efg

\section{NeedATool:  a Numerical Implementation of Needlets}
\label{sec:code}
In the previous  sections we have discussed the main needlet 
properties, which are indeed strictly related to the analytical
properties of the filter function in harmonic space, $b(\ell/B^j)$. We
now give a specific recipe for the
construction of $b(\cdot)$, and describe the main features of
``NeedATool'' Needlet Analysis Tool - a numerical code which computes the filter functions and the needlet coefficients according to the algorithm which has been applied in the several cosmological analyses \citep{Pietrobon2006ISW,Pietrobon2008AISO,Pietrobon2008NG,Pietrobon:2009qg,Cabella:2009}.

We give a step-by-step procedure, as implemented in
NeedATool and described in \cite{Marinucci2008}.

\begin{enumerate}
\item Construct the function
\[
f(t)=\left\{ 
\begin{array}{ll}
\exp (-\frac{1}{1-t^{2}})\textrm{ , } & -1\leq t\leq 1 \\ 
0, & \textrm{otherwise }
\end{array}
\right. .
\]
It is immediate to check that the function $f(\cdot)$ is $C^{\infty }$ and
compactly supported in the interval $(-1,1)$;

\item Construct the function 
\[
\psi (u)=\frac{\int_{-1}^{u}f(t)dt}{\int_{-1}^{1}f(t)dt}.
\]%
The function $\psi (\cdot)$ is again $C^{\infty };$ it is moreover
non-decreasing and normalised so that $\psi (-1)=0$ , $\psi (1)=1$;

\item Construct the function%
\[
\varphi (t)=\left\{ 
\begin{array}{lllll}
1 & \textrm{ if } & 0\leq & t & \leq \frac{1}{B} \\ 
\psi (1-\frac{2B}{B-1}(t-\frac{1}{B})) & \textrm{ if } & \frac{1}{B}\leq & t & \leq 1 \\ 
0 & \textrm{ if } & & t & >1
\end{array}
\right. 
\]
Here we are simply implementing a change of variable so that the resulting
function $\varphi (\cdot)$ is constant on $(0,B^{-1})$ and monotonically
decreasing to zero in the interval $(B^{-1},1).$ Indeed it can be checked
that
\[
1-\frac{2B}{B-1}(t-\frac{1}{B})=\left\{ 
\begin{array}{ccc}
1 & \textrm{ for } & t=\frac{1}{B} \\ 
-1 & \textrm{ for } & t=1
\end{array}
\right. 
\]
and
\begin{eqnarray*}
\varphi (\frac{1}{B}) &=&\psi (1)=1 \\
\varphi (1) &=&\psi (-1)=0\text{;}
\end{eqnarray*}
\item Construct 
\label{eq:b2_diff}
\be
b^{2}(\xi )=\varphi (\frac{\xi }{B})-\varphi (\xi ) \nonumber
\ee
The expression for $b^{2}(\cdot)$ is meant
to ensure that the function satisfies the partition-of-unity property of Eq.~\ref
{eq:partun}. Needless to say, for  $b(\xi )=\left\{ \varphi (\frac{\xi }{B}%
)-\varphi (\xi )\right\} ^{1/2}$ we take the positive root.

\end{enumerate}

Incidently, we notice that property \ref{eq:b2_diff} is crucial in allowing for the reconstruction formula, and it is shared, although within a different setup, by the implementation described by \citet{Starck2005,Starck:2009a} who makes use of the \healpix software package too. In Fig.~\ref{fig:b_filter} we show the set of filter functions in
$\ell$ space for the choice $B=2$. They result in a homogeneous binning
in $\log\ell$, whose power sum to 1, a crucial property which needlets
properties rely on.
\bfg[h]
\begin{center}
  \incgr[width=.8\columnwidth]{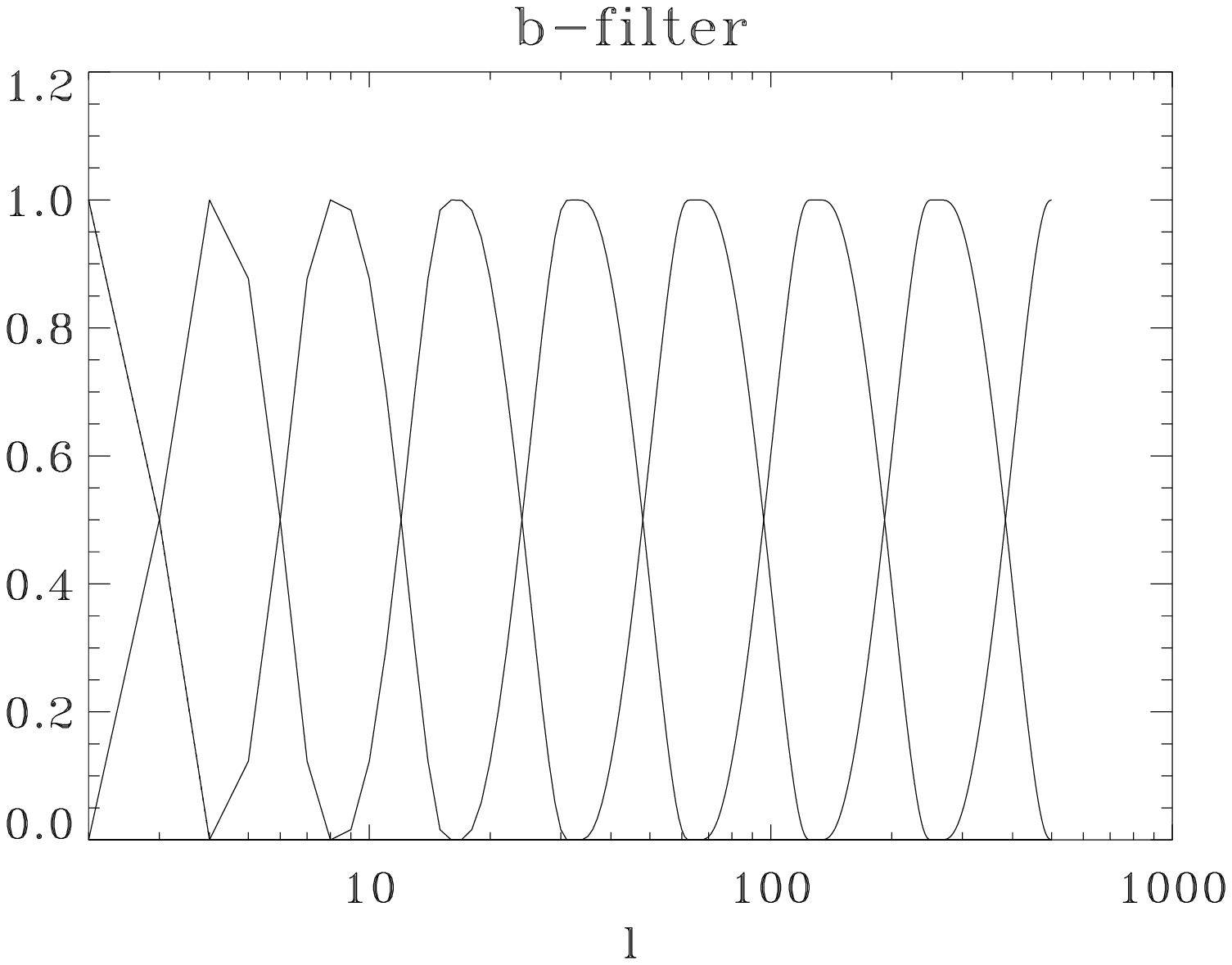}
  \caption[Filter function in $\ell$-space]{\small{Filter function in $\ell$-space which the needlet construction relies on. Set computed for $B=2$.}}
  \label{fig:b_filter}
\end{center}
\efg

NeedATool~\footnote{\href{http://www.fisica.uniroma2.it/~pietrobon/dp_files/dp_NeedATool_download.html}{http://www.fisica.uniroma2.it/$\sim$pietrobon/dp\_webpage\_eng.html}} computes the needlet filter functions in harmonic space and the needlets coefficients, given a set of parameters. The code is based on the public available \healpix package~\footnote{http://healpix.jpl.nasa.gov/}
\citep{Gorski2005Healpix}, which has to be downloaded and installed separately (version 2.10 or higher required). The software is composed of two programs ``\emph{synneed}'' and ``\emph{ananeed}''; following the \healpix structure the former deconvolves a given map to obtain a needlet frame, the latter reconstructs the original map, if a needlets basis is given as input. Both the programs, as well as the routines necessary for NeedATool will be fully integrated in the next \healpix release. Both the programs accept the same parsing file in which the fundamental parameters are provided by the user. A list of such parameters is given in Tab.~\ref{tab:need_pars}.
\begin{table*}
\center
\caption[Needlets software input parameters]{\small{Parameters required by
the codes ``synneed'' and ``ananeed''.}}
\label{tab:need_pars}
\begin{tabular}{|ccc|}
\hline
{\bf Parameter} & {\bf synneed} & {\bf Ananeed} \\
\hline
healpix\_dir & \multicolumn{2}{c|}{/usr/local/Healpix\_2.13a} \\
$\ell_{\rm max}$ & \multicolumn{2}{c|}{500} \\
$B$ & \multicolumn{2}{c|}{2.0} \\
compute\_needlets & T & \\
mapfile & input/lcdm\_map\_lmax500.fits & needlet\_2.00\_Nj009.fits\\
mapnside & 256 & 256\\
maskfile & input/sky\_cut\_1\_256\_ring.fits & input/sky\_cut\_1\_256\_ring.fits \\
bl2\_root & bl2 & \\
need\_root & needlet & recmap \\

\hline
\end{tabular}
\end{table*}
The maximum number of multipoles ($\ell_{\rm max}$) and the $B$
parameter are required. Those given, the codes computes the maximum
$j$ necessary to keep all the information in the map. The $N_{\rm
side}$ of the needlets coefficients is then determined according to
the relation $\ell_{\rm max}\leq2N_{\rm side}$. The filter functions
$b_{\ell j}$ are computed by default, while it is possible to choose
whether computing the needlet coefficients, which actually is the
most time-consuming part, by setting the keyword
``\emph{compute\_needlets}''. It is necessary then to specify the map
and its resolution. A sky mask can be applied filling the
``\emph{maskfile}'' variable. The last two keywords set the output
files. The input map of ananeed has to be the needlet output file created by synneed.

The cubature points $\mathcal{X}_{j}=\left\{ \xi_{jk}\right\}
_{k=1,2,...,}$ are  assumed to coincide with the pixelization of the unit sphere $\mathbb{S}^{2}$ provided by \healpix, with $N_{\rm side}$ such that
$l_{\rm max}\equiv[B^{j+1}]\leq 2N_{\rm side}$ (with $[\cdot]$
denoting the integer part and $B>1$). The cubature weights,
$\lambda_{\rm jk}$ are given by $1/N_{\rm pix}$, with $N_{pix}$ given
by $12\cdot N_{\rm side}^2$. We then computed the
$\beta_{jk}$ coefficients for each $k$ position given by the \healpix
scheme evaluating the projection operator, namely the product of
$\sum_{\ell\ell^\prime}Y_{\ell m}\overline Y_{ell^\prime m^\prime}$ for each pair of pixels $\xi_{jk}$, by means of the \healpix software package. The code is very fast, and it can be run on a laptop. For a low resolution map ($N_{side}=256$), it takes a few seconds, while it scales according the \healpix scaling laws for higher resolutions.

We present an example of the reconstruction power of the needlet frame. We produced a CMB map consistent with the WMAP 5-year best fit cosmological power spectrum \citep{Komatsu2008wmap5} up to $\ell_{\rm max}=500$ by using the \healpix toolbox.

We processed this map through the needlet pipeline extracting first the needlet coefficients by applying \emph{synneed}, then reconstructing the map using \emph{ananeed}. We repeated this procedure both for the whole sky case and in presence of symmetric sky cuts, $15^\circ$, $5^\circ$, $1^\circ$ and $0.1^\circ$. The results are shown in Fig.~\ref{fig:recmap} (left column) together with the percentage error due to the procedure (right column). The error due to the forward and backward transformation is smaller then $0.01\%$ except for few pixels in the whole sky case. The result worsens reaching $~1\%$ when a broad mask is applied, basically because the needlet coefficients at very large scale are affected by the presence of the mask. This effect is actually expected, since needlets work well at small angular scales, namely when the filter function $b(\ell/B^j)$ groups several multipoles. The mask effect may be reduced by either fine tuning the $B$ parameter, or deconvolving the mask effect . Reconstruction error is indeed wavelet frame dependent: a better performance can be achieved when the wavelet is defined as the difference between two scaling functions at two different resolutions instead of the difference between the square root of the scaling functions. This is the implementation discussed by \cite{Starck2005} which leads to an exact reconstruction.

\begin{figure*}[!h]
\center
\incgr[width=.5\columnwidth, angle=90]{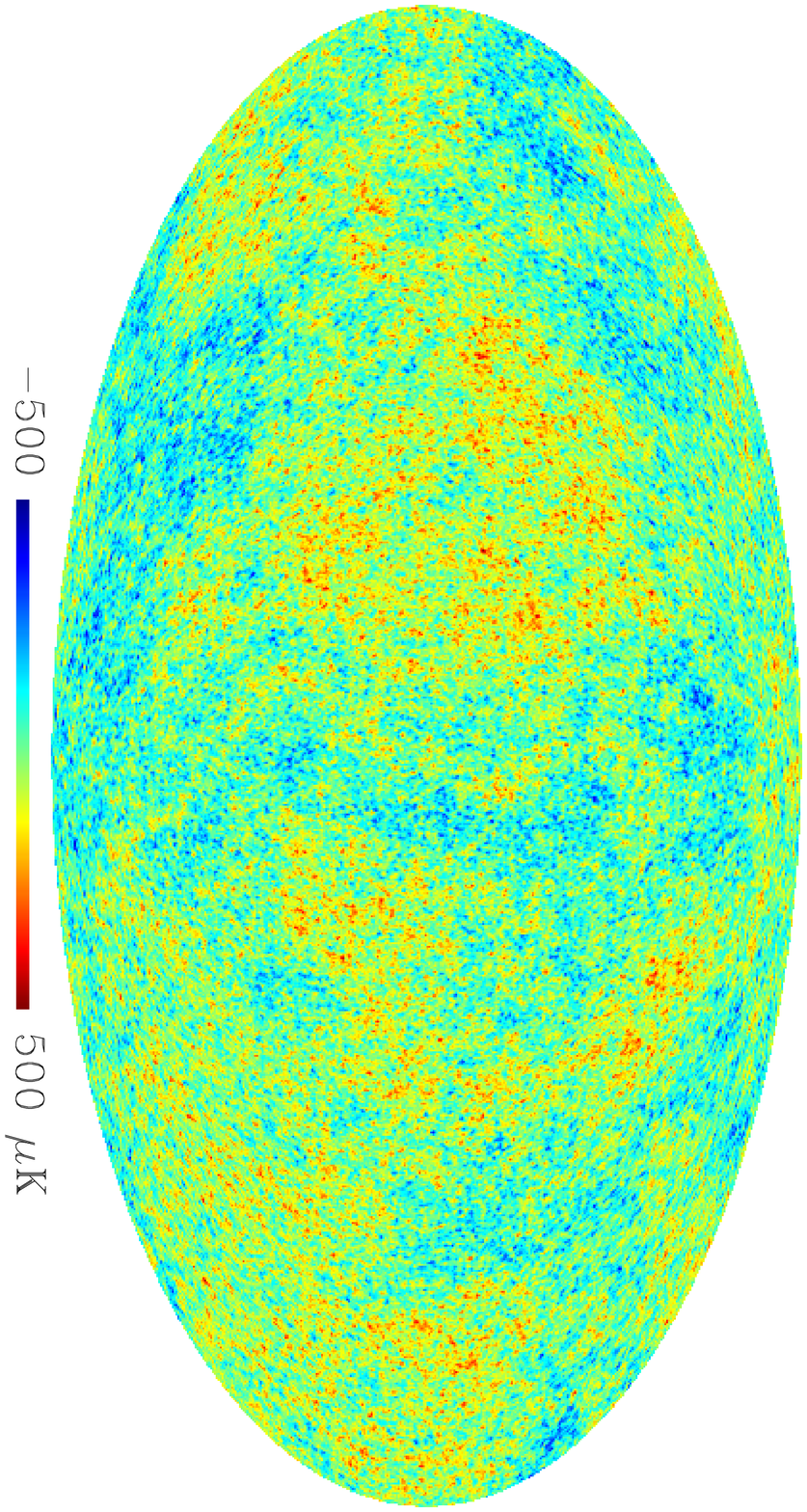}
\incgr[width=.5\columnwidth, angle=90]{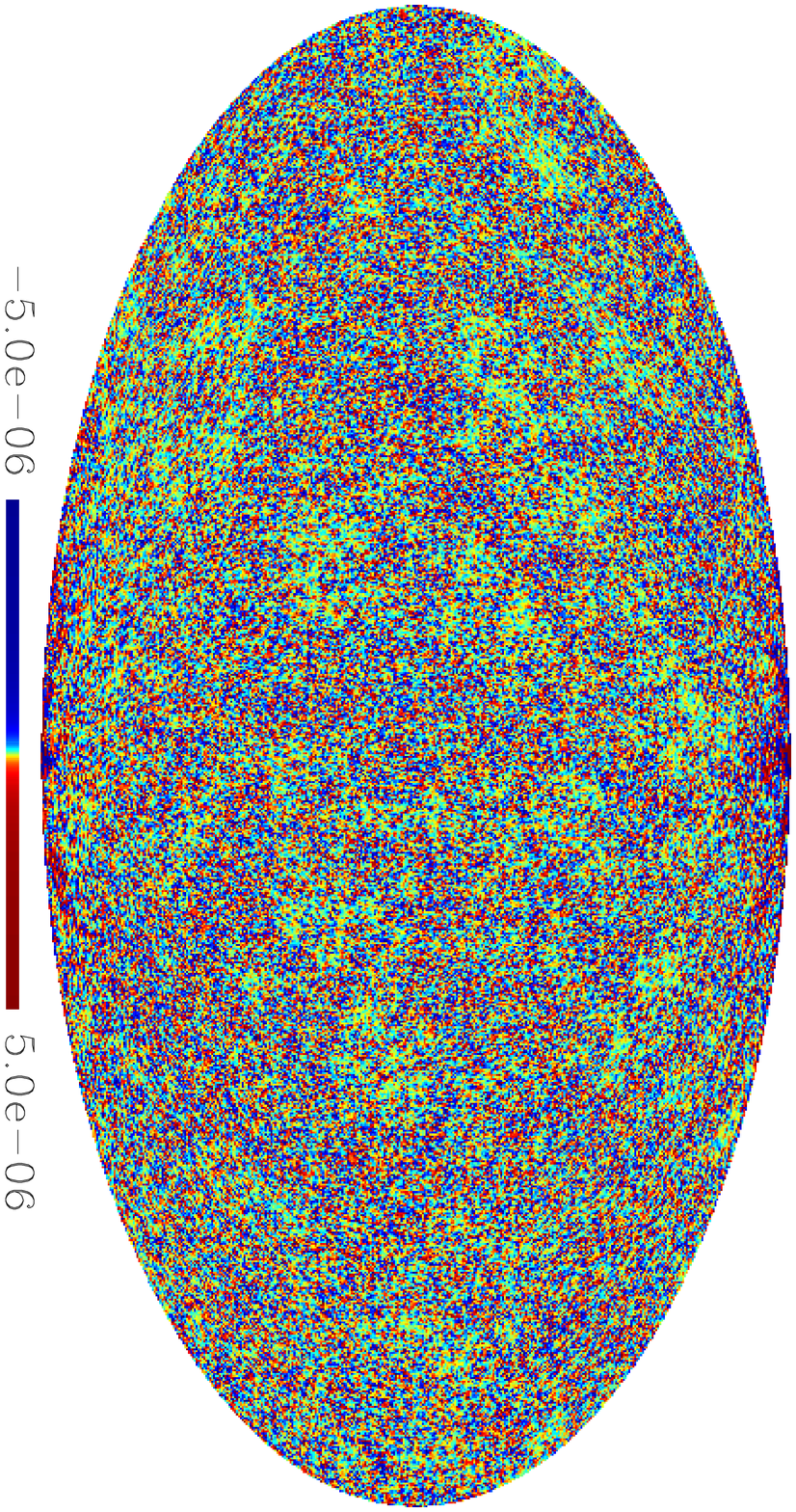}
\incgr[width=.5\columnwidth, angle=90]{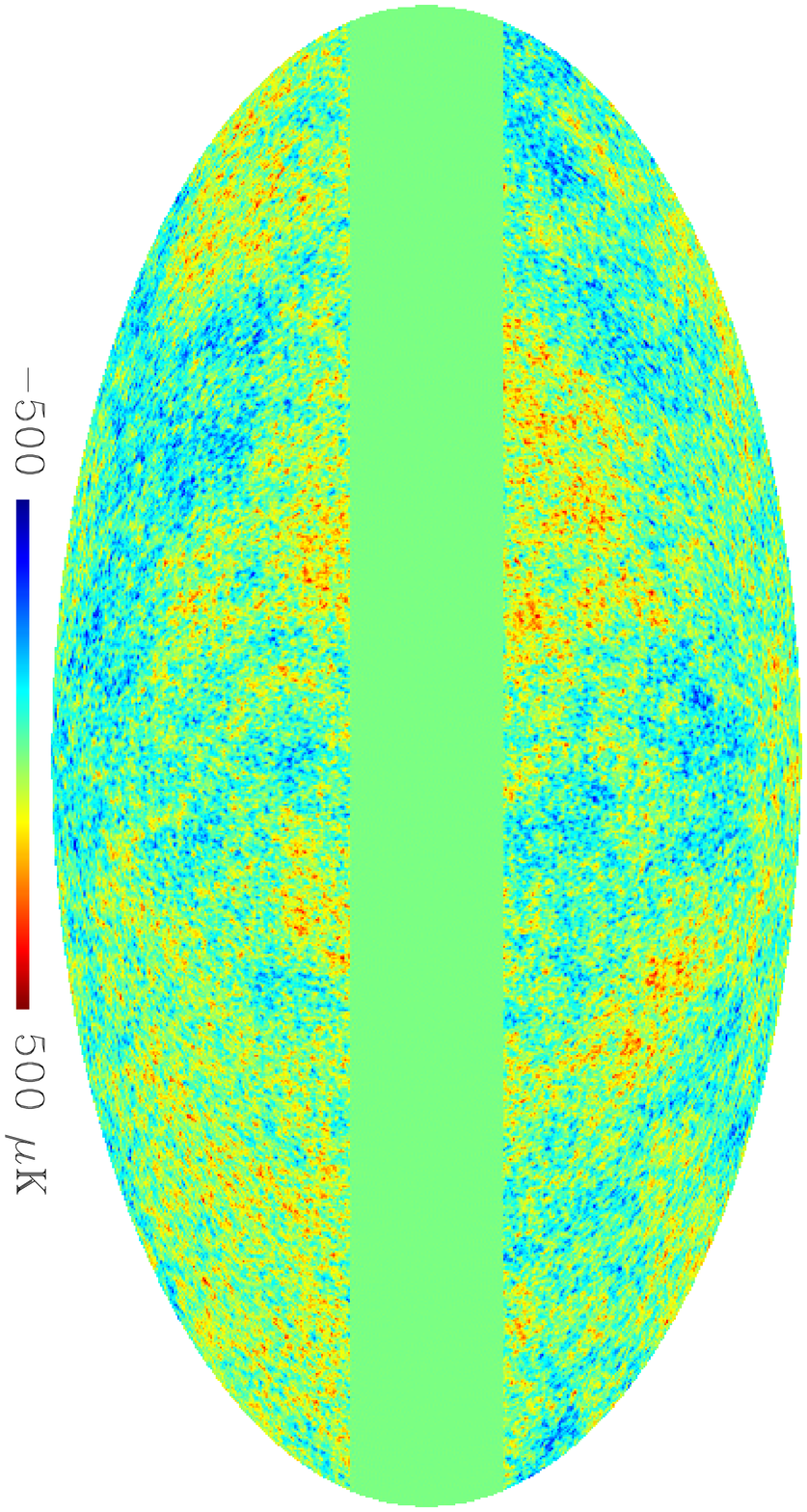}
\incgr[width=.5\columnwidth, angle=90]{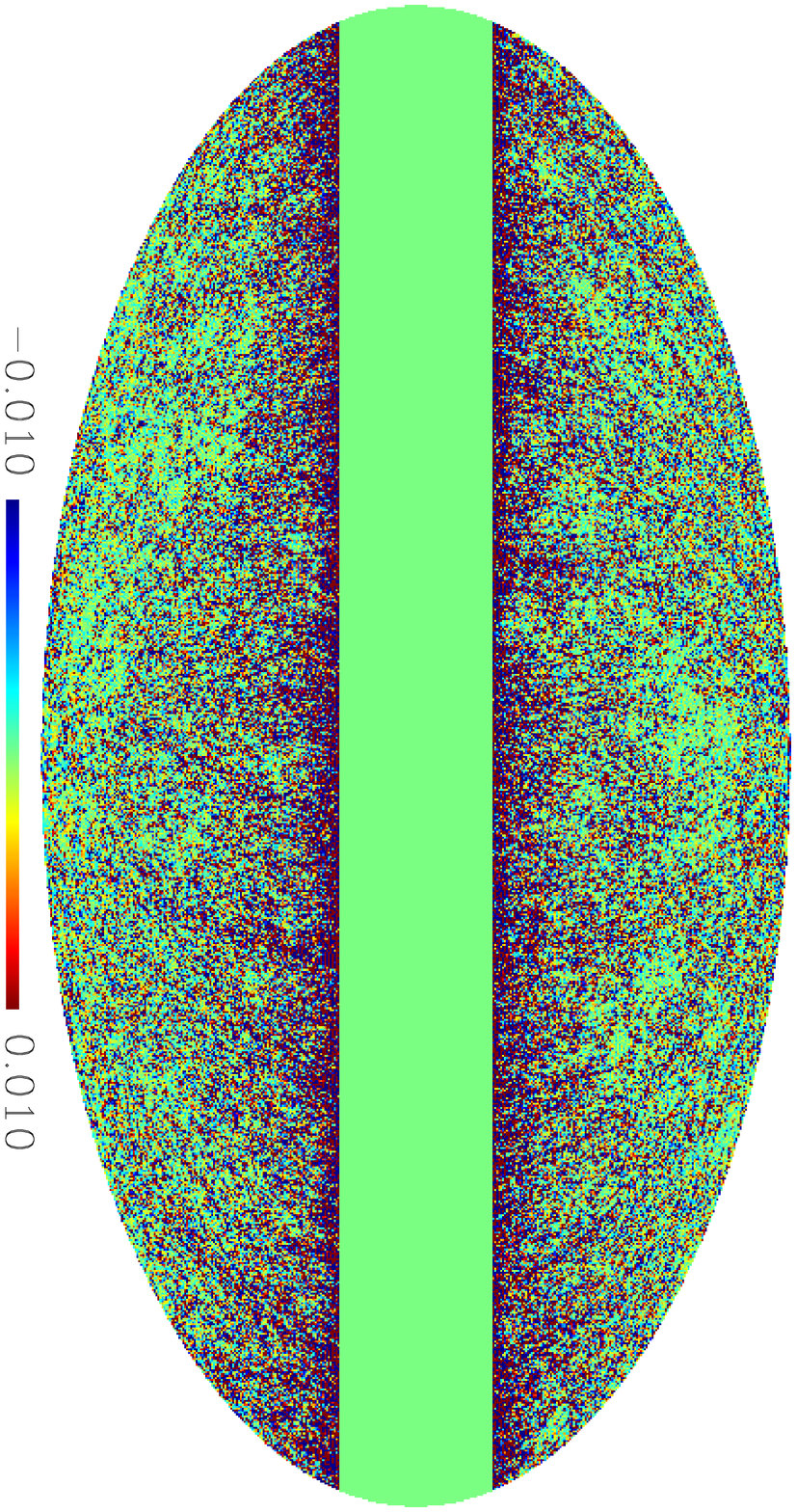}
\incgr[width=.5\columnwidth, angle=90]{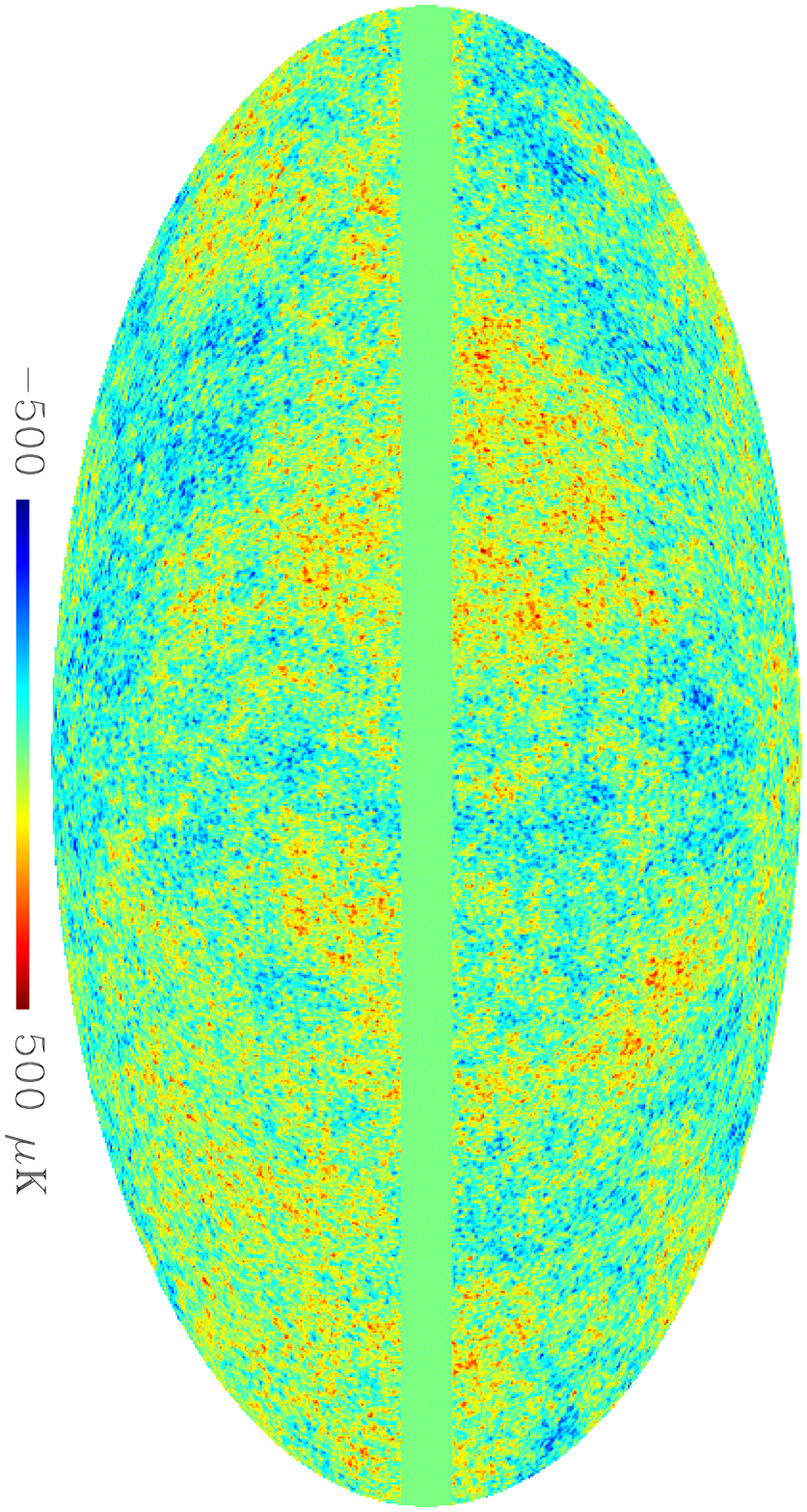}
\incgr[width=.5\columnwidth, angle=90]{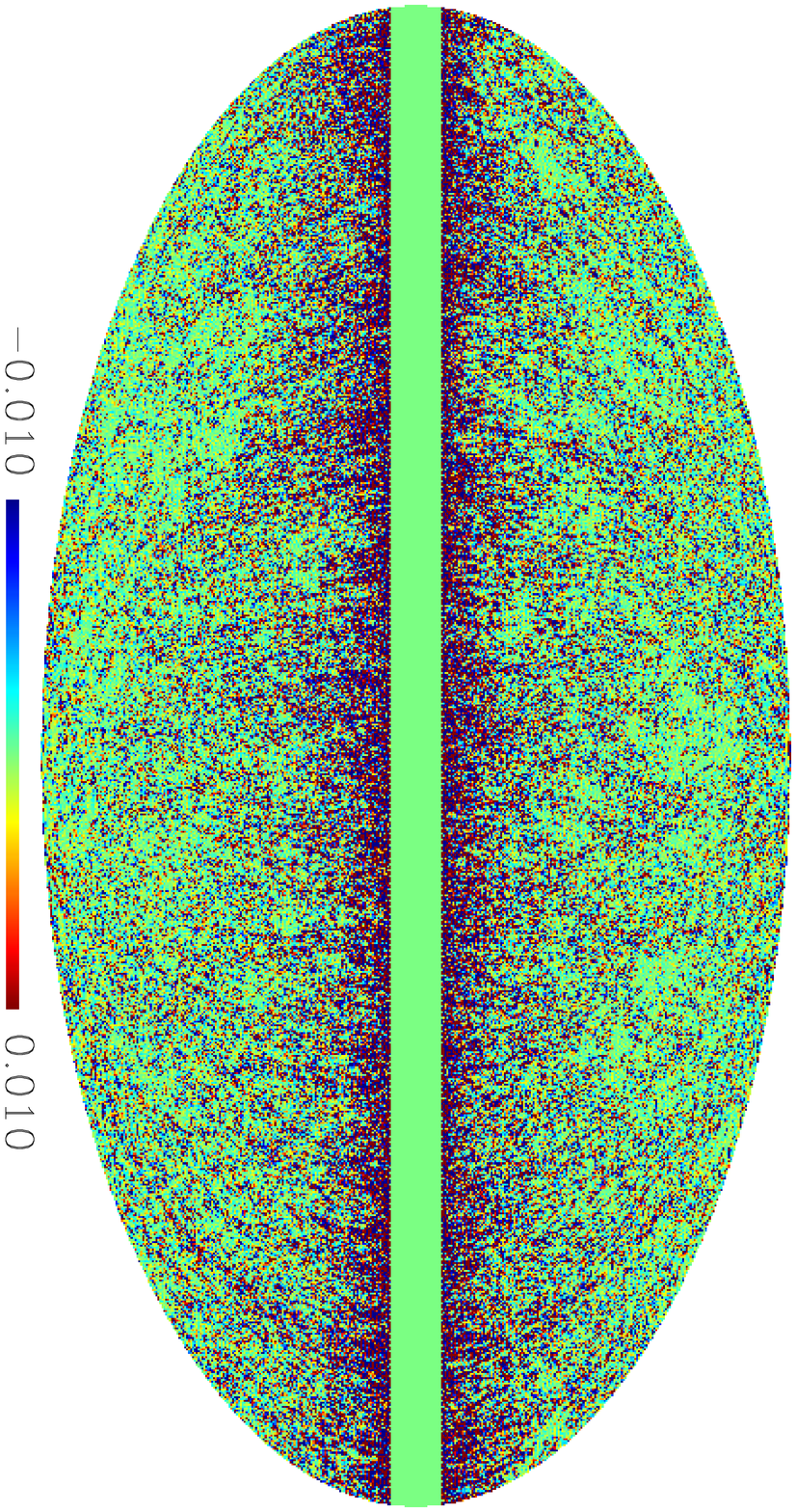}
\incgr[width=.5\columnwidth, angle=90]{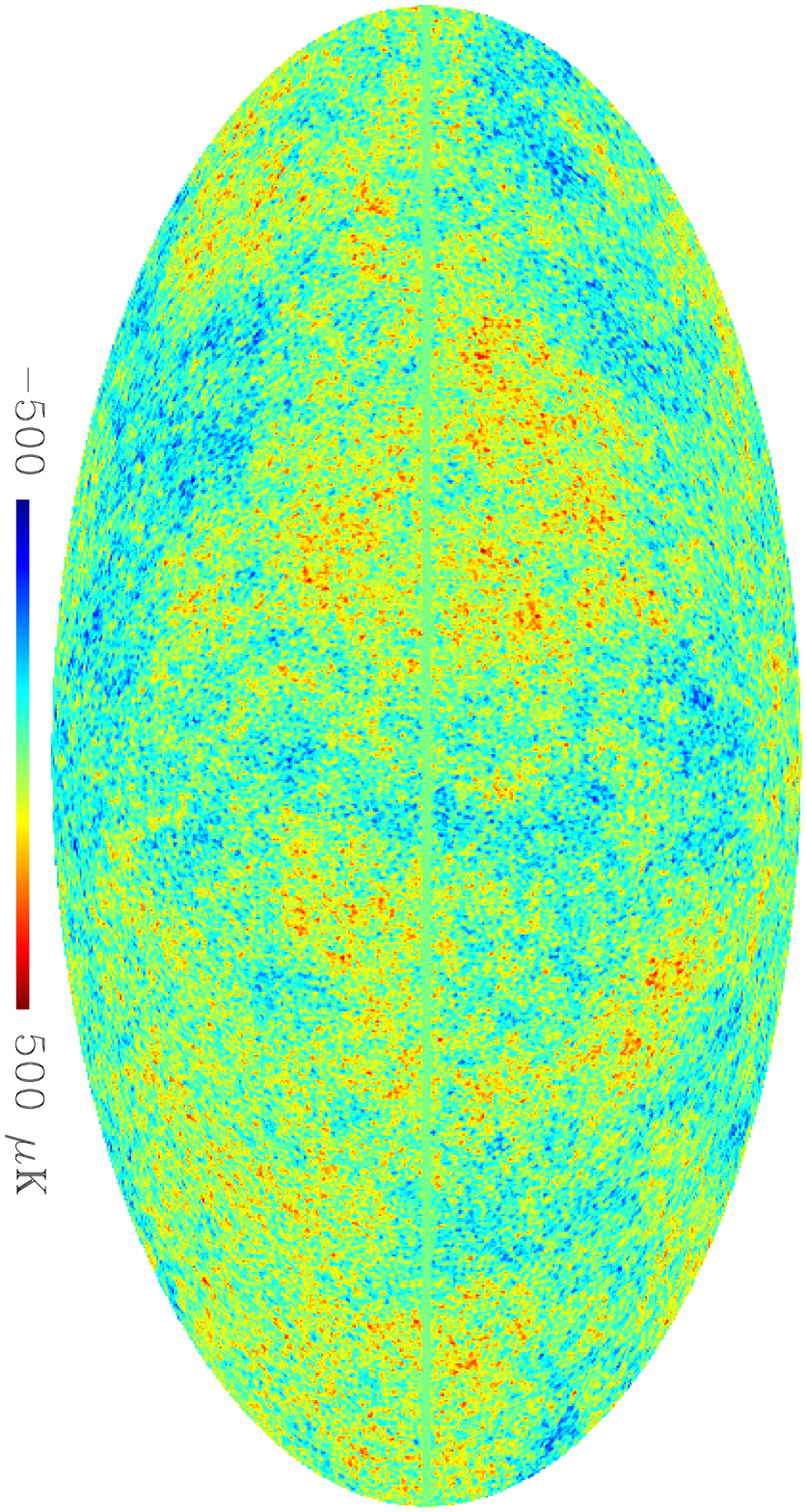}
\incgr[width=.5\columnwidth, angle=90]{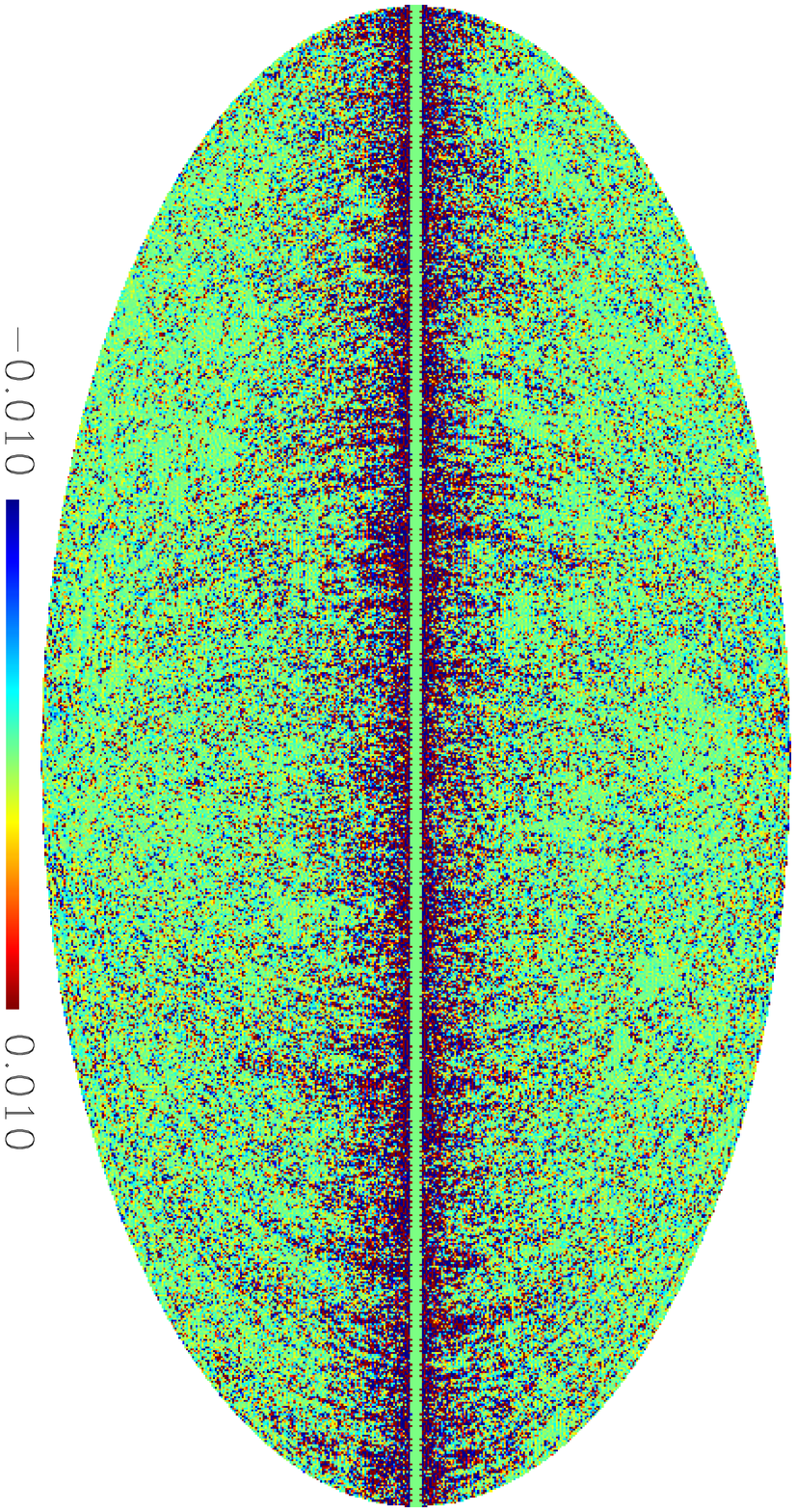}
\caption{Left: reconstructed maps for the several sky cuts analysed (from top to bottom full-sky, 15$^\circ$, 5$^\circ$ and 1$^\circ$); right: percentage error on the reconstructed map.}
\label{fig:recmap}
\end{figure*}

To give another estimate of the error due to the reconstruction, we computed the angular power spectrum of the original map, $\mathcal{C}_\ell$, as well those of the reprocessed ones, $\mathcal{C}_\ell^{\rm R}$, and compared them. The ratio is displayed in Fig.~\ref{fig:spectra} (left panel) for the four analysed sky cuts applied. The full sky reconstruction shows an excellent agreement, the difference between the power spectra being of the order of $10^{-4}$; when a mask is present, the reconstruction causes an error of few percent at very low angular scales, which decreases at small scales.
\begin{figure*}
\center
\incgr[width=.3\textwidth, angle=90]{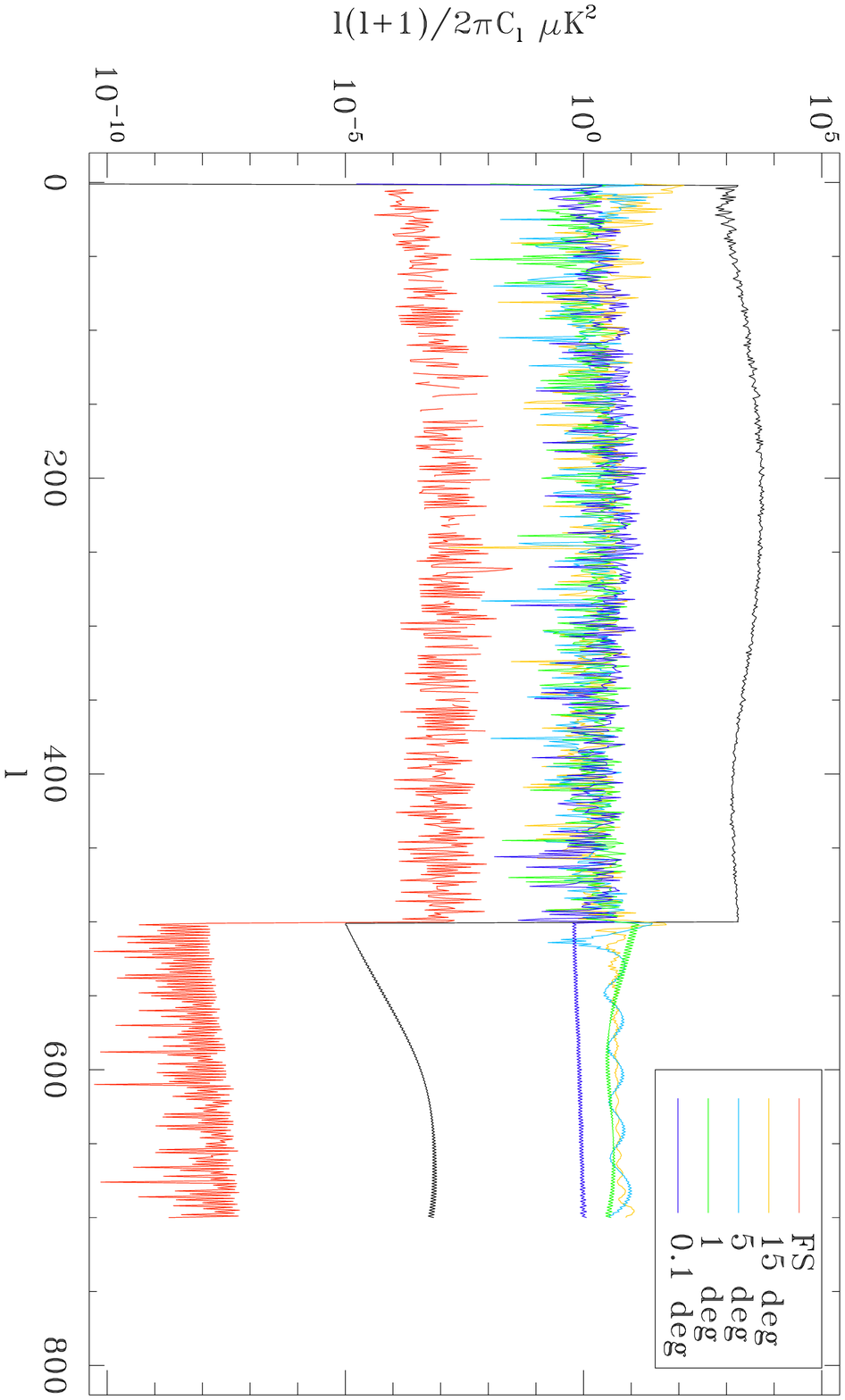}
\incgr[width=.3\textwidth, angle=90]{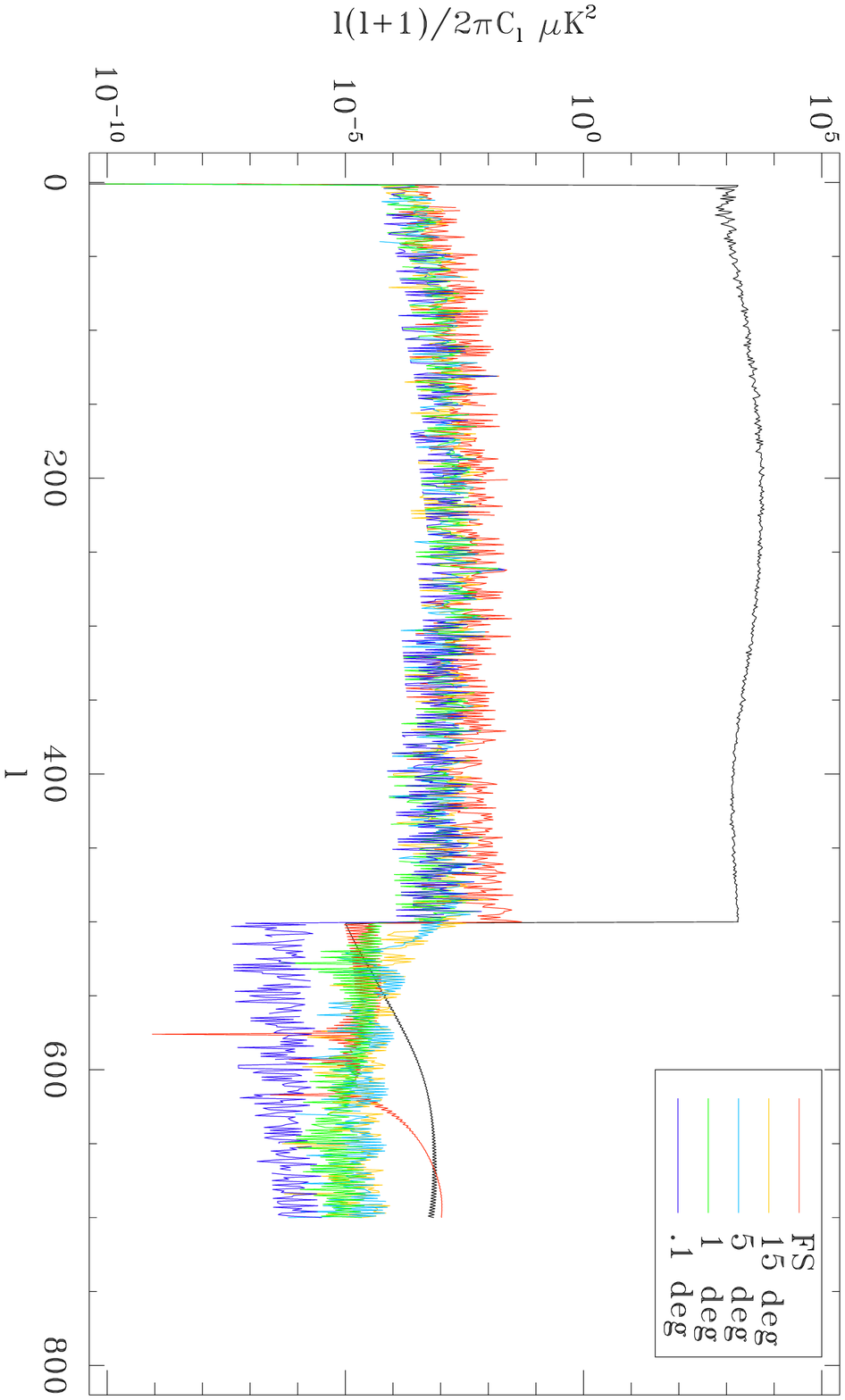}
\caption{Left panel: absolute value of the difference between the power spectra extracted from the original map and the reconstructed one for several symmetric sky cuts: full sky (red solid line), 15$^\circ$ (yellow solid line), 5$^\circ$ (light blue solid line), 1$^\circ$ (green solid line) and 0.1$^\circ$ (blue solid line). The black solid line shows the CMB angular power spectrum for the original map. Right panel: comparison between the reconstruction using {\it synfast-anafast} combination and the needlets pipeline for the same sky cuts. The input angular power spectrum contains information up to $\ell=500$, but we computed the power spectrum of the realization map up to $\ell=700$ to check the leakeage from low multipoles to high ones: this explains the edge at $\ell=500$ in the plotted power spectra.}
\label{fig:spectra}
\end{figure*}

The non perfect reconstruction in presence of missing information is expected. Therefore, it is interesting to compare the needlets performance to the one of the spherical harmonics implemented by \healpix. We extracted the $a_{\ell m}$ from the same CMB realisation for the five different sky coverages considered by using {\it anafast}; then we produced a CMB map out of the {\it pseudo}-$a_{\ell m}$ using {\it synfast}. We finally analysed the resulting map computing its power spectrum. The difference with respect the power spectrum computed through needlets is shown in the right panel of Fig.~\ref{fig:spectra}. The spectra agree very well regardless the applied mask. This confirms the high performance of needlets in reconstructing a field in the sphere. As a further figure of merit, we show in Fig.~\ref{fig:map15_rec} the percentage error between the two maps reconstructed using spherical harmonics and needlets for the broadest mask (15$^\circ$). The agreement is striking: a large scale pattern appears at the level of $<1\%$.
\bfg[h]
\begin{center}
\incgr[width=.6\columnwidth, angle=90]{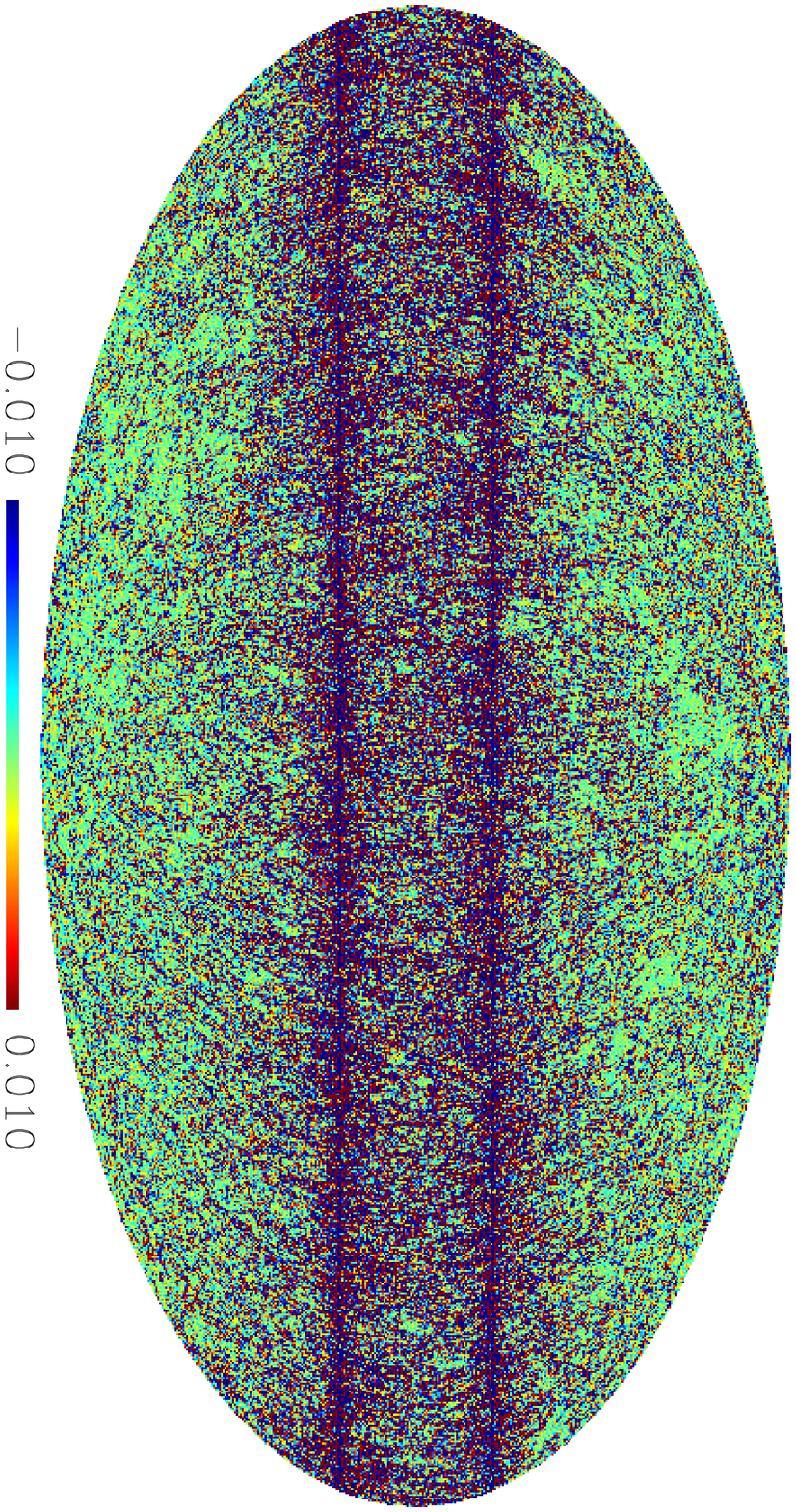}
\caption{Percentage error between the two maps reconstructed using spherical harmonics and needlets. A symmetric sky cut of 15$^\circ$ is applied. The agreement is impressive.}
\label{fig:map15_rec}
\end{center}
\efg

We further investigate the mask effect, focusing on how the presence of a sky cut couples the spherical harmonics coefficients. As it is well known, in the case of partial sky coverage the spherical harmonics do not form an orthonormal basis. This can be formalized as follows:
\bea
&&T^R(\hg) = \sum_{\ell m}\tilde{a}_{\ell m}Y_{\ell m}(\hg) \\
&& \tilde{a}_{\ell^\prime m^\prime} = \int_{\mathcal{O}}T(\hat\gamma)\overline{Y}_{\ell^\prime m^\prime}\de\Omega = \sum_{\ell m}a_{\ell m}\mathcal{K}_{\ell\ell^\prime m m^\prime} \\
&& \int_{\Omega}\overline{Y}_{\ell^\prime m^\prime}(\hat\gamma)Y_{\ell m}(\hat\gamma)W(\hat\gamma)\de\Omega \equiv \mathcal{K}_{\ell\ell^\prime m m^\prime},
\label{eq:kern}
\eea
where $\mathcal{O}$ is the observed region and $\mathcal{K}_{\ell\ell^\prime m m^\prime}$ is the coupling matrix. In the needlet framework, this affects the $\beta_{\rm jk}$ coefficients as:
\bea
&&\tilde{\beta} _{\rm jk} = \sqrt{\lambda_{\rm jk}}\sum_{\ell^\prime m^\prime} b\Big(\frac{\ell^\prime}{B^j} \Big) \tilde{a}_{\ell^\prime m^\prime} Y_{\ell^\prime m^\prime}(\xi_{\rm jk})\\
&&T^{\rm R}(\hg) = \sum_{\rm j} \sum_{\ell^\prime m^\prime} b\Big(\frac{\ell^\prime}{B^j} \Big) \tilde{a}_{\ell^\prime m^\prime}\sum_{\ell m} b\Big(\frac{\ell}{B^{j}}\Big)Y_{\ell m}(\hat\gamma) \mathcal{K}_{\ell\ell^\prime m m^\prime}. \nonumber \\
\label{eq:need_rec}
\eea
As for the spherical harmonics decomposition, the map reconstructed through the needlet pipeline is computed from $\tilde{a}_{\ell m}$; in addition, the mask shows indirectly its effect on the filter functions $b(\ell/B^j)$.

To visualize this effect, we plot in Fig.~\ref{fig:kernels} the coupling matrix $\mathcal{K}_{\ell\ell^\prime m m^\prime}$ for a full sky case (upper panel), in which it actually becomes a Kronecker's $\delta$ function, $\delta_{\ell\ell^\prime mm^\prime}$; for a 15$^\circ$ symmetric sky cut (middle panel) and for the convolution $\sum_jb(\ell^\prime/B^j)\mathcal{K}_{\ell\ell^\prime mm^\prime}b(\ell^\prime/B^j)$, which appears in Eq.~\ref{eq:need_rec}. Since the numerical evaluation of the coupling matrix up to $\ell=500$ is a severe computational challenge and we expect the higher multipoles to be marginally affected by a partial sky coverage, we computed Eq.~\ref{eq:kern} on a smaller range of spherical harmonics in the interval $\ell\in[0,20]$ on the \healpix pixelization at $N_{\rm side}=16$. The presence of the mask translates into off-diagonal terms in Fig.~\ref{fig:kernels}: each multipole is coupled to its second neighbor, whereas the coupling with the first neighbor is vanishing because of parity. The coupling is indeed not negligible up to the fourth second neighbor. The lower panel in Fig.~\ref{fig:kernels} clearly depicts the properties of needlets and explain the error we obtain in the reconstruction. The correlation matrix results a superposition of blocks, each of them corresponding to a given j: each block partially overlaps only the first neighbor, as a consequence of the compact support of the $b(\ell/B^j)$ functions. (This is shown by the white regions we observe in the plot). Moreover, the off-diagonal elements are less powerful compared to the auto-correlation terms and the correlation extends to a lower number of multipoles. At very large scale however, a lack of power is present in the diagonal terms which affects the global reconstructed map. This can be seen in the right panel of Fig.~\ref{fig:recmap}, where the error in the reconstructed map follows a peculiar pattern which results symmetric as a consequence of the sky cut.

The additional coupling due to the mask present in the needlet construction arises from the definition of needlet itself, Eq.~\ref{eq:needlets_expansion}, since the needlet is built from the projection operator. However, needlets provide a natural way to counterbalance this effect. At large scale, the needlets are not sharply localised, and the coefficients turn of to be non-vanishing also in the masked region. Taking into account this leakage, and including in the reconstruction procedure the coefficients inside the mask, it is possible to reduce the effect of the coupling between needlets. This is the actual procedure implemented in the code, which leads to the results discussed in this section. The residual effect due to the presence of the mask shows up in the large scale pattern highlighted in Fig.~\ref{fig:map15_rec}. 
\begin{figure*}
\center
\incgr[width=.3\textwidth]{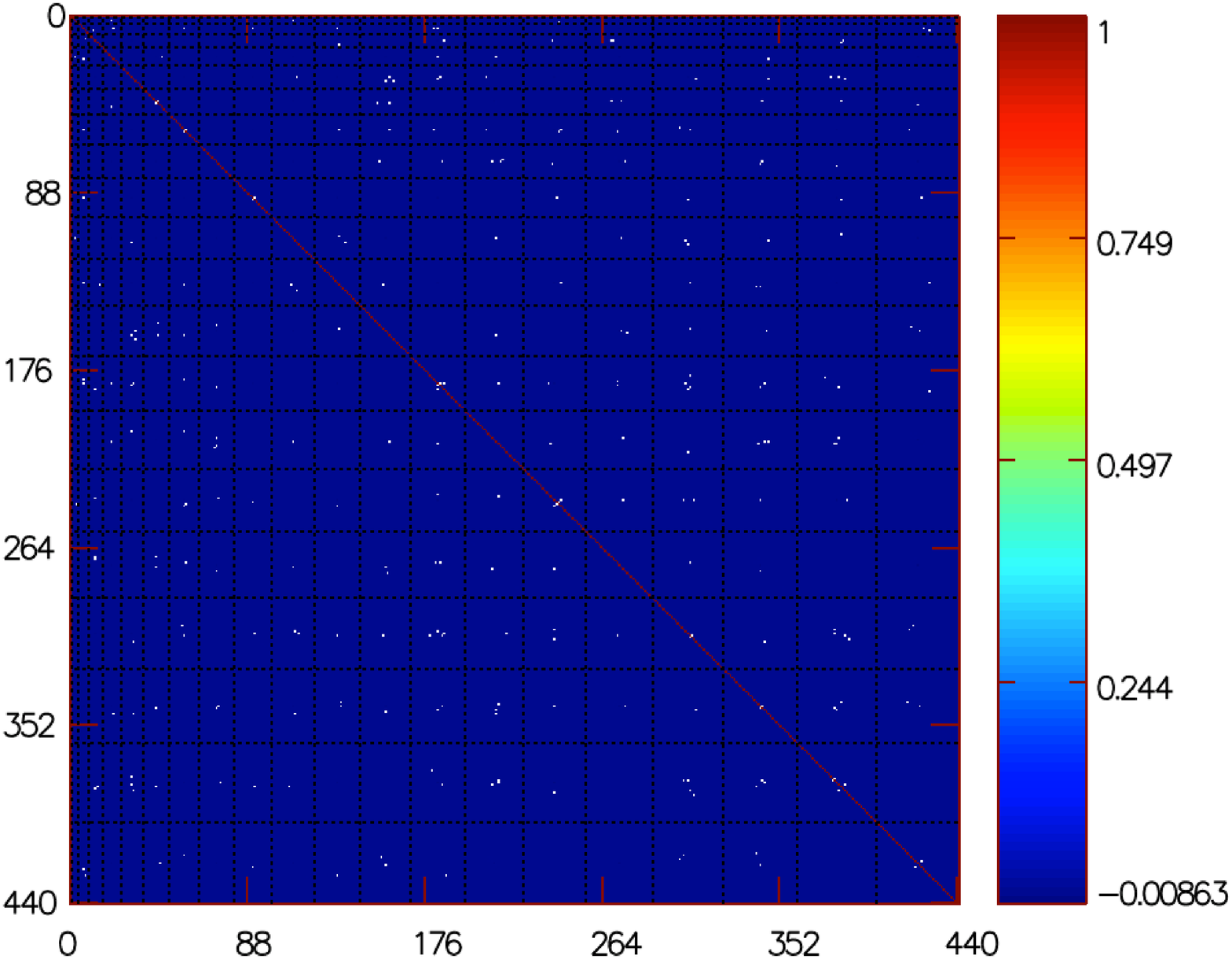}
\incgr[width=.3\textwidth]{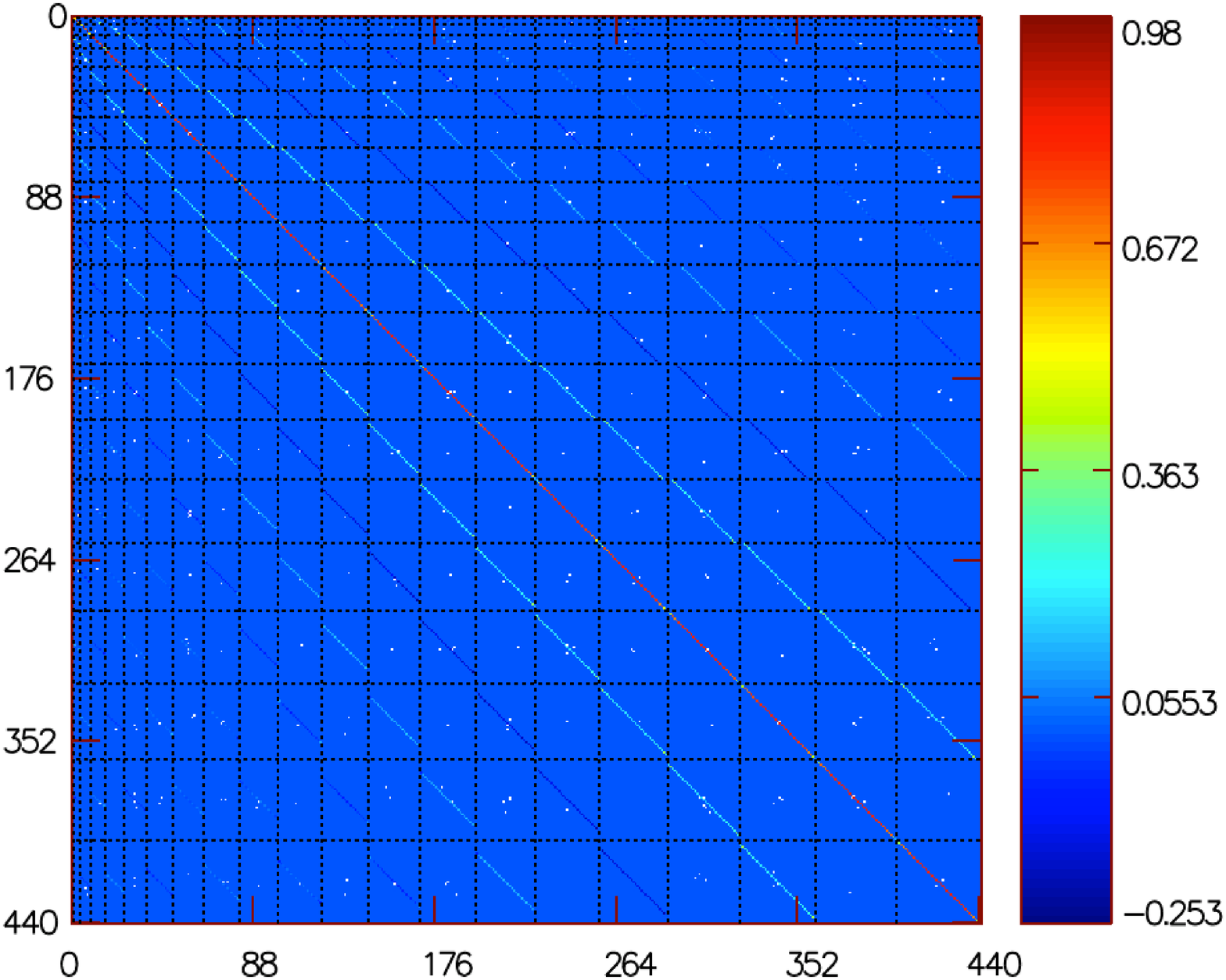}
\incgr[width=.3\textwidth]{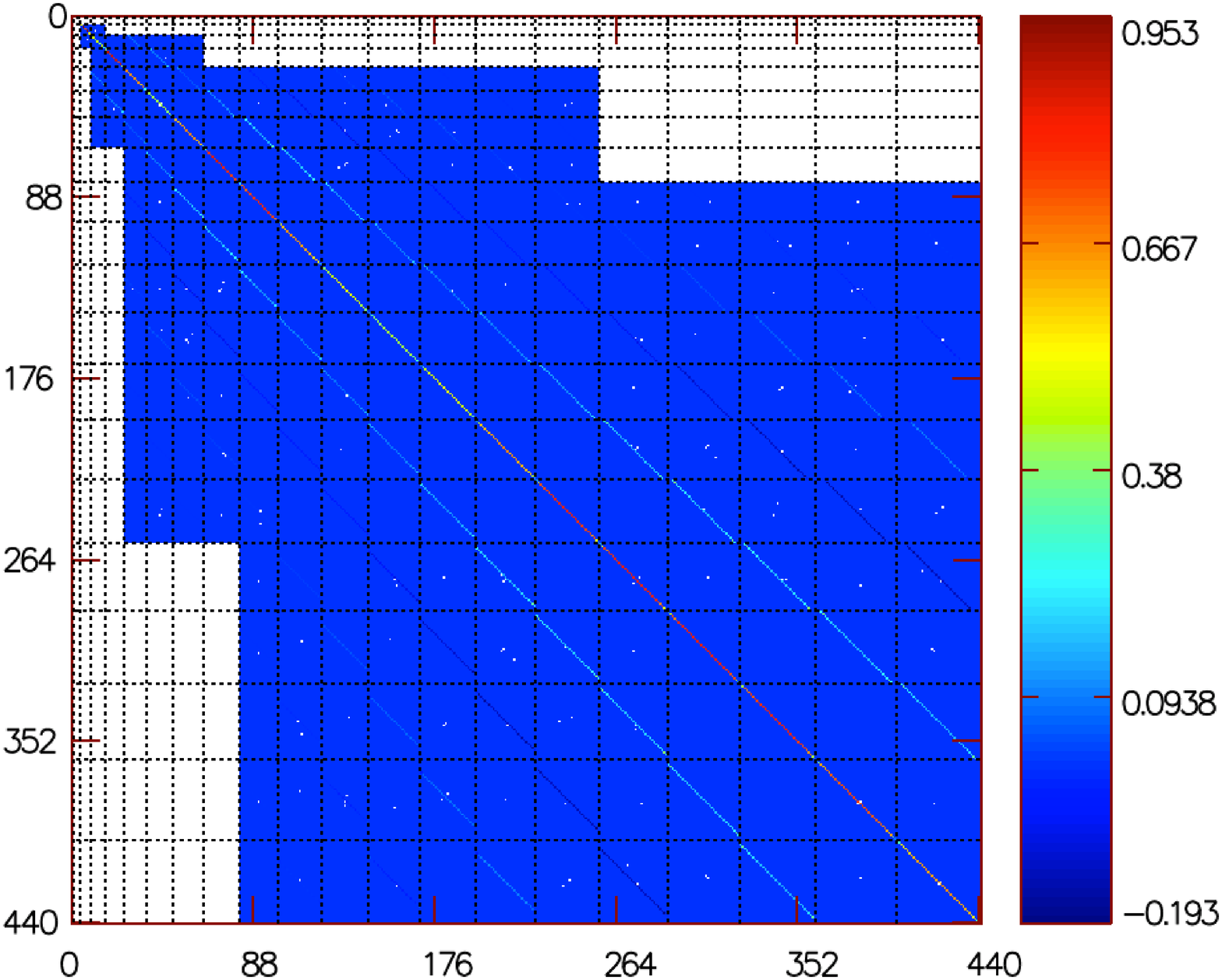}
\caption{Coupling matrix as a function of the quantity $\ell^2+\ell+m+1$ for three interesting cases: full sky (left panel), 15$^\circ$ symmetric sky cut (middle panel) and its convolution with needlet window functions (right panel). The grid marks the multipoles.}
\label{fig:kernels}
\end{figure*}


\section{Conclusions}
\label{sec:conclusions}
In this paper we have introduced NeedATool, a public software for the analysis of datasets on the sphere based on the needlet framework. The software is particularly useful for the analysis of CMB data, as shown by its successful application to, e.g., the WMAP dataset.  The needlet construction differs from other wavelet renditions due to the distinctive properties of the filter functions $b_{\ell j}$, which translate into a sharp localisation in pixel space and excellent properties of non-correlation among the functions of the set. This aspect is crucial when building estimators for CMB data analysis as we discussed extensively: therefore, needlets are a very promising tool for high-accuracy cosmological experiments. 

\section*{Acknowledgements}
We thank Domenico Marinucci for useful discussions and Giancarlo de Gasperis and Rajib Saha for technical support.

\bibliography{Biblio}





\end{document}